\documentclass[published]{JHEP3} % 10pt is ignored!
\newcommand\ee{\end{equation}}

\JHEP{00(2011)000}

\JHEPspecialurl{http://jhep.sissa.it/JOURNAL/JHEP3.tar.gz}

\usepackage{amssymb,epsfig,multicol,multicol,bbm}

\newcommand\fverb{\setbox\fverbbox=\hbox\bgroup\verb}
\newcommand\fverbdo{\egroup\medskip\noindent%
                        \fbox{\unhbox\fverbbox}\ }
\newcommand\fverbit{\egroup\item[\fbox{\unhbox\fverbbox}]}
\newbox\fverbbox

\usepackage{amsmath}

\usepackage{epsfig}

\def\p{\partial}                                % partial derivative
\def\slashchar#1{\setbox0=\hbox{$#1$}                   % set a box for #1
   \dimen0=\wd0                                         % and get its size
   \setbox1=\hbox{/} \dimen1=\wd1                       % get size of /
   \ifdim\dimen0>\dimen1                                % #1 is bigger
      \rlap{\hbox to \dimen0{\hfil/\hfil}}              % so center / in box
      #1                                                % and print #1
   \else                                                % / is bigger
      \rlap{\hbox to \dimen1{\hfil$#1$\hfil}}           % so center #1
      /                                                 % and print /
   \fi}

\newcommand{\beq}{\begin{equation}}
\newcommand{\eeq}{\end{equation}}
\newcommand\be{\begin{equation} }
\newcommand\bea{\begin{eqnarray}}
\newcommand\eea{\end{eqnarray}}

\def\endtitle{\par\end{quotation}\vskip3.5in minus2.3in\newpage}

     \def\x{\hat x}

       \def\P{\Pi}

  \title
  {\bf  \Large  Noncommutative $\kappa$-Minkowski $\phi^4$ theory:\\ 
 Construction, properties and propagation }

\author{S. Meljanac$^1$, A. Samsarov$^1$, J. Trampeti\'c$^{1,2}$ and M. Wohlgenannt$^3$ \\ 
1. Rudjer Bo\v skovi\' c Institute, 
P.O.Box 180, HR-10002 Zagreb, Croatia \\
2. Max-Planck-Institut f\"ur Physik, (Werner-Heisenberg-Institut),
  	 F\"ohringer Ring 6, D-80805 M\"unchen, Germany\\
3. Faculty of Physics, University of Vienna, Boltzmanngasse 5, A-1090 Vienna, Austria\\
E-mail: \email{meljanac@irb.hr}, \email{asamsarov@irb.hr}, \email{josipt@rex.irb.hr }
\email{michael.wohlgenannt@univie.ac.at}}

\received{\today}               %%
\accepted{\today}               %% These are for published papers.

\abstract{ Noncommutative (NC) $\kappa$-deformation of a spacetime, 
whose NC coordinates close in a Lie algebra, 
affects the coalgebra of the Poincar\' e group 
and the algebra of physical fields. This leads to a
modification of the multiplication in 
the corresponding universal enveloping algebra. The usual pointwise multiplication has to be 
replaced by a deformed star product. 
The measure problems in the $\kappa$-Minkowski spacetime 
are avoided because the measure function is naturally absorbed 
in the new ${\star}_h$-product. 
That reflects itself in the way we have constructed
the deformed NC scalar $\phi^4$ action. 
The action is further modified by adding  
a harmonic oscillator term and expanding up to linear order in 
the deformation parameter $a$. This leads to an effective theory on commutative spacetime.
Furthermore, we obtain modified equations of motion 
and conserved currents.  
In order to compute the tadpole contributions 
to the scalar field propagation/self-energy, we anticipate
that statistics on $\kappa$-Minkowski spacetime is specifically deformed.  
Thus, our prescription in fact corresponds to a {\it hybrid approach}
of standard quantum field theory (QFT) and NCQFT on 
$\kappa$-deformed Minkowski spacetime.
The results are analyzed in 
the framework of the two-point Green's function
for low, intermediate and Planckian energies, respectively.
For low energies $E$, the dependence of the tadpole contribution 
on the $\kappa$-deformation parameter $a$ completely drops out.
At Planckian propagation energies, it tends to 
a finite fixed value, which only depends on the Planckian energy 
and/or the $\kappa$-deformation parameter. 
{\it Semiclassical/hybrid behavior} of the first order quantum effects 
{\it do show up} due to the $\kappa$-deformed momentum conservation law. 
The mass term of the scalar field is shifted and these shifts are 
dramatically different at different propagation energies. 
At Planckian energies, the $\kappa$-modified dispersion relations 
{\it show the genuine effect of birefringence}.
We conclude that our results could have physical consequences 
e.g. in the NC Higgs sector and the connection to quantum gravity.}


\begin{document}
%\maketitle
 
%\newpage

%%%%%%%%%%%%%%%%%%%%%%%%%%%%%%%%%%%%%%%%%%%%%%%%%%%%%%%%%%%%%%%%%%%
%%%%%%%%%%%%%%%%  Introduction %%%%%%%%%%%%%%%%%%%%%%%%%%%%%%%%%%%%%this model
%%%%%%%%%%%%%%%%%%%%%%%%%%%%%%%%%%%%%%%%%%%%%%%%%%%%%%%%%%%%%%%%%%%

\section{Introduction}

Two basic areas of interest, namely quantum gravity and the deformation of the Poincar\'{e} algebra, 
make $\kappa$-Minkowski spacetime to
an important subject of theoretical investigations from both,
physical as well as mathematical perspective. 

An argument in favor of $\kappa$-Minkowski spacetime 
is the indication that it could arise in the context of
quantum gravity coupled to matter fields \cite{AmelinoCamelia:2003xp,Freidel:2003sp}. 
These considerations
show that after integrating out the gravitational topological degrees of freedom of
gravity, the effective dynamics of matter fields is described by a
noncommutative quantum field theory which has a $\kappa$-Poincar\'{e}
group as its symmetry \cite{Freidel:2005bb,Freidel:2005me,Freidel:2005ec}. 
In this context,
the $\kappa$-Minkowski spacetime can be, and there are some arguments to
support this, considered as a flat limit of quantum gravity.

Second, $\kappa$-Minkowski spacetime emerges naturally from 
the $\kappa$-Poincar\'{e} algebra 
\cite{Lukierski:1991pn,Lukierski:1992dt,Majid:1994cy,Zakrzewski:1994,Lukierski:1993wx},
which provides a possible group theoretical framework for the 
describing %symmetry lying in the core 
of the so-called Doubly Special Relativity (DSR) theories \cite{AmelinoCamelia:2000ge,AmelinoCamelia:2000mn,
Magueijo:2001cr,Magueijo,KowalskiGlikman:2002we,KowalskiGlikman:2002jr}.
Although different proposals for DSR theories can be considered as different
bases \cite{KowalskiGlikman:2002we,KowalskiGlikman:2002jr}
for the $\kappa$-Poincar\'{e} algebra, they all represent the same noncommutative structure.
This situation makes noncommutative field theories on noncommutative 
$\kappa$-Minkowski spacetime an even more interesting
subject to study. Various attempts have been undertaken in this
direction by many authors
\cite{Kosinski:1999ix,Kosinski:1999dw,AmelinoCamelia:2001fd,
Daszkiewicz:2004xy,Dimitrijevic:2003wv,Ghosh:2006cb,Ghosh:2007ai}.
%,
%including various possibilities for construction 
%and investigation of their properties.

Recently, it was established that $\kappa$-Minkowski spacetime leads to
modification of the particle statistics described by deformed oscillator algebras
\cite{Govindarajan:2009wt,Young:2007ag,Daszkiewicz:2007az,Arzano:2007ef,Daszkiewicz:2007ru,
Daszkiewicz:2008bm,Arzano:2008bt}.
Deformation quantization of Poincar\'{e} algebra can be 
performed by means of the twist operator
\cite{drinfeld,drinfeldN,Borowiec:2004xj,Borowiec:2006fc,Balachandran:2007vx} 
which happens to
include the dilatation generator. Thus, the twist belongs is an element of the universal
enveloping algebra of the general linear algebra
\cite{Bu:2006dm,Govindarajan:2008qa,Borowiec:2008uj,Bu:2009tc,Kim:2009jk}.
It also gives rise to a deformed statistics on $\kappa$-Minkowski spacetime 
\cite{Young:2007ag,Govindarajan:2008qa,Young:2008zm,Young:2008zg}.
However, since the $\kappa$-Poincar\'{e} Hopf algebra is a quantum symmetry
described only approximatively by the twisted quantum algebra,  
identifying charges and conserved current and therefore also establishing a Noether theorem could be problematic. 
A certain modification of momentum conservation
is necessary in order to obtain a reasonable physics out of the twisted Hopf algebra prescription of our theory. 

The transition from the $\kappa$-deformed to the commutative Minkowski spacetime 
in the case of the free fields was described in \cite{Freidel:2006gc},
while the star product and interacting fields were treated in the same approach in \cite{KowalskiGlikman:2009zu}.
The problem of UV/IR mixing and $\kappa$-deformation was discussed in 
\cite{Grosse:2005iz}.
In correspondence to the above observations, 
as well as by comparing the deformed dispersion
relations to the corresponding time delay calculations of high energy photons,
bounds can be put on the deformation (quantum gravity) scale 
\cite{Harikumar:2009wv,Arzano:2009bd,Borowiec:2009ty}.
In the present work, we are work with commutative Minkowski spacetime endowed with a $\kappa$-deformed star product.
The considered $\kappa$-deformed symmetry of the Minkowski spacetime has an undeformed Lorentz sector 
(classical basis \cite{Kosinski:1994br,KowalskiGlikman:2002we,KowalskiGlikman:2002jr,
Borowiec:2009vb}) and its noncommutative 
coordinates close in the $\kappa$-deformed Lie algebra and additionally, form a Lie
algebra together with the Lorentz generators
\cite{Meljanac:2006ui,Meljanac:2007xb,KresicJuric:2007nh,Meljanac:2010ps}.
This deformation of the spacetime
structure affects the algebra of physical fields, leading to a
modification of the multiplication in the corresponding universal
enveloping algebra. As mentioned before, the usual pointwise multiplication has to be replaced by a
deformed star product, i.e. by the new star product ${\star}_h$. 
This star product together with the usual integration enjoys the important trace property \cite{Meljanac:2010ps}.
%The action for a massive, complex scalar field in interaction
%gets modified accordingly.
The integral measure problems for $\kappa$-Minkowski spacetime 
are avoided, since the measure function is naturally absorbed 
in the new ${\star}_h$-product.  
The action is further modified by adding a 
harmonic oscillator term \cite{Grosse:2005da,Grosse:2004yu} 
and expanding up to first order in the deformation parameter $a$. The resulting action should be considered as 
an effective theory on commutative spacetime.
Furthermore, we obtain modified equations of motion 
and conserved currents to first order in $a$.
Next, in order to compute the tadpole contributions 
to the scalar field propagation/self-energy, we anticipate
that statistics on $\kappa$-Minkowski spacetime is specifically deformed. 
Expanding the model was necessary in order to be able 
to compute relevant physical quantities, 
such as the self-energy of the complex scalar field $\phi$. 
The above properties are very welcome, however, 
we have to stress that by truncating the $\kappa$-deformed action
at the linear order in the deformation parameter $a$ 
we have lost nonperturbative quantum effects 
like the celebrated UV/IR mixing \cite{Grosse:2005iz}, 
which, for example, connects NC field theories 
to Holography via UV and IR cutoffs 
in a model independent way \cite{Horvat:2010km,Cohen:1998zx}.
Resummation of the expanded action can in principle restore the 
nonperturbative character of the theory, see e.g. 
\cite{Horvat:2011iv,arXiv:1109.2485,arXiv:1111.4951}. 
%Under such circumstances UV/IR mixing might help
%to determine what the UV theory might be, that is, it would help to
%determine UV completeness of the theory. 
Those are general properties of most of the 
NCQFT expanded/resummed in terms of 
the noncommutative deformation parameter. 
Holography and %the NCGFT 
UV/IR mixing are %in the literature known as
possible windows to quantum gravity 
\cite{Horvat:2010km,Cohen:1998zx,Szabo:2009tn}.
  
In the same line of reasoning,
we should take into account the harmonic oscillator term, despite that it is well
known that such a term breaks translational invariance \cite{Grosse:2005da,Grosse:2004yu}.
Our approach generally represents a {\it hybrid approach} of
standard quantum field theory and NCQFT on 
$\kappa$-Minkowski spacetime, including $\kappa$-deformed momentum conservation law
\cite{Kosinski:1999ix,AmelinoCamelia:2001fd,
Daszkiewicz:2004xy,KowalskiGlikman:2004qa,AmelinoCamelia:2001me}.

In a next step, we discuss the results in the framework of two-point Green's function
for low, intermediate and the Planck scale energy regime, respectively.
We have found semiclassical behavior of the first order quantum effects,
and, as a consequence, a shift of 
the mass term of the scalar field. This shift depends on the energy.
Thus, the dispersion relations are $\kappa$-deformed and 
we have found a genuine birefringence effects, \cite{Abel:2006wj,Buric:2010wd}
of the massive scalar field mode at first order in the deformation parameter $a$. 
This is similar to the fermion field
birefringence in truncated Moyal $\star$-product theories \cite{Buric:2010wd}.  

Above, we have described the main results of this paper,
which could be of physical importance for example
for the $\kappa$-deformed scalar field (Higgs) and its propagation,
as well as to quantum gravity.

In the first section, we give some mathematical preliminaries including the
Hopf algebra structure of $\kappa$-deformed Minkowski spacetime 
and star products. In the second section, we introduce,
the hermitian realization and the corresponding star product $\star_h$.
The modified $\kappa$-deformed scalar field action based on the above notions 
is introduced next, and the equations of motion are derived. The corresponding currents 
are conserved. The properties of our new $\kappa$-deformed action 
%like the interpretation of the $\kappa$-deformed action in terms of 
%the deformed theory (with undeformed fields) 
on ordinary Minkowski spacetime are discussed in the last section.
This includes $\kappa$-deformed Feynman rules and the field propagation 
via computation of two-point Green's function within the proposed model.

\section{Mathematical preliminaries of $\kappa$-deformed Minkowski spacetime}
\subsection{Hopf algebra and star product}

We are considering a $\kappa$-deformation of the Minkowski spacetime whose
symmetry has an undeformed Lorentz sector and whose noncommutative
coordinates $\x_{\mu}, ~ (\mu = 0,1,...,n-1),$ 
close in a Lie algebra together with the Lorentz generators
$M_{\mu\nu},~ (M_{\mu \nu} = -M_{\nu \mu}) $ \,:
\begin{align} 
\label{2.1}
 [\x_{\mu},\x_{\nu}] & = i(a_{\mu}\x_{\nu}-a_{\nu}\x_{\mu})\,, \\ 
 [M_{\mu \nu}, M_{\lambda \rho}]  & =  \eta_{\nu \lambda}M_{\mu \rho} -
 \eta_{\mu \lambda}M_{\nu \rho}
 -\eta_{\nu \rho} M_{\mu \lambda} + \eta_{\mu \rho} M_{\nu \lambda}\,, 
 \label{2.2} \\
 [M_{\mu\nu}, \x_{\lambda}] & = \x_{\mu} \eta_{\nu\lambda} - \x_{\nu}
  \eta_{\mu\lambda}-i\left( a_{\mu}  M_{\nu\lambda}-a_{\nu} 
  M_{\mu\lambda} \right)\,, 
  \label{2.3}
\end{align}
where the deformation parameter $a_{\mu}$ is a constant
Lorentz vector and $~ \eta_{\mu\nu} = diag(-1,1,\cdot \cdot \cdot, 1)$
defines the metric in this spacetime. The quantity $a^2 = a_{\mu}a^{\mu}$ 
is Lorentz invariant having a dimension of inverse mass squared,
 $a^2 \equiv \frac{1}{{\kappa}^{2}}.$ 
  The above algebra satisfies all the Jacobi identities
 and therefore forms a Lie algebra with the property that in the
 limit $a_{\mu} \rightarrow 0,$ the commutative spacetime  with the
 usual action of the Lorentz algebra is recovered. 
 Throughout the paper we
will work in natural units,  $\hbar = c =1$. 

The symmetry of the deformed spacetime (\ref{2.1}) is assumed to be
described by an undeformed Poincar\'{e}
algebra. Thus, in addition to Lorentz generators  $M_{\mu\nu}, $ we
also introduce the momentum $p_{\mu}$ which transforms as a vector under the
Lorentz algebra, 
\begin{align} \label{2.4}
[p_{\mu},p_{\nu}]&=0\,,  \\
[M_{\mu\nu},p_{\lambda}]&= \eta_{\nu\lambda}\,
p_{\mu}-\eta_{\mu\lambda}\, p_{\nu}\,. 
\label{2.5}
\end{align}
For convenience we refer to algebra (\ref{2.1})-(\ref{2.5}) as a
deformed special relativity algebra since its different
realizations lead to different special relativity models with
different physics encoded in deformed dispersion relations resulting
from such theories. This algebra, however, does not fix the commutation
relation between $~p_{\mu}$ and $\x_{\nu}$. In fact, there are
infinitely many possibilities 
%for the commutation
%relation between $~p_{\mu}$ and $\x_{\nu},$ all of which 
which are all consistent
with the algebra (\ref{2.1})-(\ref{2.5}), in the sense that the Jacobi
identities between all generators of the algebra are satisfied.   
In this way, we have an extended algebra generated by $M_{\mu\nu}, ~p_{\mu}$ and $\x_{\lambda}$.
%and has Jacobi identities satisfied for all combinations of the generators.
Particularly, the algebra generated by $p_{\mu}$ and $\x_{\nu}$ is a deformed
Heisenberg-Weyl algebra that can generally be written in the form
\begin{equation} 
\label{2.6}
[p_{\mu},\x_{\nu}] = -i \Phi_{\mu\nu}(p)\,,
\end{equation}
where $ \Phi_{\mu\nu}(p)$ are functions of the generators $p_{\mu}$,
which are constrained by the Jacobi identities 
and the boundary conditions
$ \Phi_{\mu\nu}(0)=\eta_{\mu\nu}$. The latter constraint reflects 
the requirement that the deformed NC space reduces to
ordinary commutative space in the limiting case of vanishing
deformation parameter, $a_{\mu} \rightarrow 0.$ 

The momentum $p_{\mu} = -i \hat{\p}_{\mu}$, expressed in terms
of the deformed derivative $ \hat{\p}_{\mu}$, can be realized in a
natural way \cite{Meljanac:2007xb} by adopting the identification between deformed and
undeformed derivatives, $ \hat{\p}_{\mu} \equiv \p_{\mu}$, implying
$p_{\mu} = -i \p_{\mu}$.
The deformed algebra (\ref{2.1})-(\ref{2.5}) then admits a wide class of realizations
\begin{equation} 
\label{2.7}
\x_{\mu} =x^\alpha \Phi_{\alpha\mu}(p)\,,
\end{equation}
\begin{equation} 
\label{2.8}
M_{\mu\nu} = x_{\mu}\p_{\nu}- x_{\nu}\p_{\mu}\,,
\end{equation}
in terms of the undeformed Heisenberg algebra
\begin{equation} 
\label{2.9}
[x_{\mu},x_{\nu}]=0, \qquad [\p_{\mu},\p_{\nu}]=0, \qquad [\p_{\mu},x_{\nu}]=\eta_{\mu\nu}
\end{equation}
and the analytic function $\Phi_{\alpha\mu}(p)$
satisfying the boundary conditions
$~ \Phi_{\alpha\mu}(0)=\eta_{\alpha\mu}$. At this point $\partial_{\mu}$'s 
represent abstract generators of the undeformed Heisenberg algebra.
Later on, we will use a concrete representation in terms of differential
operators, $\partial/\partial x^{\mu}$.
By taking this prescription, the deformed algebra
(\ref{2.1})-(\ref{2.5}) is then automatically satisfied, as
well as all Jacobi identities among the generators $\x_\mu$, $M_{\mu\nu}$, and $p_{\mu}$. 

Some comments are in order.
The algebra (\ref{2.1})-(\ref{2.5}), as it is defined, does not closed. 
Closing it by the relations
(\ref{2.6}), we obtain an extended algebra containing the deformed Heisenberg-Weyl
subalgebra. The realization (\ref{2.7}) indicates that the deformed Heisenberg-Weyl algebra (\ref{2.1}),
(\ref{2.4}), (\ref{2.6}) and the undeformed one (\ref{2.9}) 
are isomorphic at the level of vector spaces.
In the same way one can show that the deformed and the undeformed extended algebras
are isomorphic in the same sense. This is independent of the particular choice for $\Phi_{\mu \nu}$ in (\ref{2.7}), 
as shown in a general setting in \cite{Borowiec:2010yw}, where the extended algebra (\ref{2.1})-(\ref{2.5}) 
has been defined as a crossed (smash) product algebra. In Ref \cite{Borowiec:2010yw}, the authors started from the coproduct (Hopf algebra + module algebra) and
then determined the crossed commutation relations of 
the extended algebra (\ref{2.1})-(\ref{2.6}). Our
route is just opposite: we close the algebra by the crossed commutation
relation and then accordingly determine the coproduct.

In this paper, we shall strictly work with
one particular type of realization (as well as with its hermitian
variant, see Eq.(\ref{2.10d}) below), 
namely with the realization of the form \cite{KresicJuric:2007nh} 
\begin{equation} 
\label{2.10}
\x_{\mu}=
 x_{\mu} \sqrt{1+ a^2 p^2 } - iM_{\mu \nu} a^{\nu}\,.
\end{equation}
We refer to this type of realization as Maggiore-type of realization 
of the algebra (\ref{2.1}), since it leads to phase space noncommutativity
analyzed for the first time by Maggiore \cite{Maggiore:1993rv,Maggiore:1993zu}. This
particular kind of phase space noncommutativity 
is directly related to a generalized uncertainty  principle appearing in
the contexts of string theory and quantum gravity. 
It is also considered in \cite{Dimitrijevic:2003wv}.
The realization (\ref{2.10}) belongs to the class defined in (\ref{2.7}) and is thus consistent with algebra (\ref{2.1})-(\ref{2.5}).
With these particular settings, the deformed Heisenberg-Weyl
algebra (\ref{2.6}) receives the following form 
\begin{equation} 
[p_{\mu},\x_{\nu}]=-i\eta_{\mu\nu} \left(ap+\sqrt{1+ a^2 p^2}
  \right) +i a_\mu p_\nu\,.
\label{2.11}
\end{equation}

As mentioned above, this particular type of phase space noncommutativity leads
to uncertainty relations of the form 
\cite{Maggiore:1993rv,Maggiore:1993zu,Giddings:2007bw}
\begin{equation}
\triangle x_{\mu} ~ 
\ge ~ \frac{\hbar}{\triangle p_{\mu}}+\alpha G \triangle p_{\mu}\,,
\label{2.12}
\end{equation}
($\alpha$ is constant, and $G$ is the gravitational constant) 
that have been obtained from the study of string collisions at
Planckian energies, i.e. so called gravity collapse of strings \cite{Giddings:2007bw}, 
thus manifesting its dynamical origin.
The same generalized uncertainty principle emerges from considerations
related to quantum gravity \cite{Maggiore:1993rv,Maggiore:1993zu}.

The operator appearing in (\ref{2.11}) in parenthesis appears to play
 a very important role in $\kappa$-deformed spaces in
 general. Therefore, we give a special label to it,
\begin{equation} 
\label{16}
 Z^{-1} = ap + \sqrt{1+ a^2 p^2}\,.
\end{equation}
This operator, among others, has the following  properties that define 
a universal shift operator \cite{KresicJuric:2007nh}:
\begin{equation} 
\label{2.13}
[Z^{-1},\x_{\mu}] = -ia_{\mu} Z^{-1}, \qquad [Z,p_{\mu}] =0\,,
\end{equation}
with $Z$ being its inverse, $Z = \frac{1}{Z^{-1}}$.
The operator $Z = \frac{1}{Z^{-1}}$ and its  properties  
are described in great detail in Refs. \cite{KresicJuric:2007nh,Meljanac:2010ps}.

It can certainly be expected that a deformation of the spacetime
structure will affect the algebra of physical fields, leading to a
modification of the multiplication in the corresponding universal
enveloping algebra. Specifically it means that a spacetime deformation
requires one to replace the usual pointwise multiplication by a
deformed product or star product, which will finally have its consequences in physics,
particularly it will reflect itself in the way in which one should construct
the field theoretic action.
 This modified multiplication, i.e. star product will
 obviously depend on the particular realization (\ref{2.7}), 
 which in turn is characterized
 by an analytic function $\Phi_{\mu \nu}$ of generators
 $p_{\mu}$. In this paper it will be
understood that noncommutative coordinates  $\x,$
appearing in all subsequent course of exposition, will be represented
either by the specific realization determined by (\ref{2.7}) or by its
hermitian variant, (\ref{2.10d}) in the next section.

In preparation for our further analysis it is useful to introduce a few notions.
First, we denote by ${\mathcal{A}}$ the algebra
of physical fields $\phi(x)$ in commutative coordinates
$x_{\mu}$. Since the the polynomials in $x_{\mu}$ build a basis for the physical fields $\phi(x)$, the algebra  ${\mathcal{A}}$ is the universal enveloping algebra generated by the commuting
coordinates $x_{\mu}$. The algebra  ${\mathcal{A}}$ can also be understood as a module 
of the deformed Weyl algebra, which is generated by
$\x_{\mu}$ and $\p_{\mu}, ~~ \mu = 0,1,...,n-1$, and allows for 
infinite series in $\p_{\mu}$.  In a similar way as for commutative fields $\phi(x)$,
the noncommutative fields $ \hat{\phi}(\hat{x})$ are built of polynomials in $\hat{x}_{\mu}$,
and thus belong to the universal enveloping algebra
${\hat{\mathcal{A}}}_{\kappa}$ generated by the noncommutative coordinates $\hat{x}_{\mu}$.
Although the universal enveloping algebras ${\mathcal{A}}$
and ${\hat{\mathcal{A}}}_{\kappa}$ are not isomorphic to each other,
there exists a
unique map and even an isomorphism  at the level of vector
 spaces for any given
realization described by the function $\Phi_{\mu \nu}$ in 
(\ref{2.7}).
Both enveloping algebras, ${\mathcal{A}}$
and ${\hat{\mathcal{A}}}_{\kappa}$, can be shown to
have a Hopf algebra structure \cite{Lukierski:1991pn,Majid:1994cy,majid1}, 
with the latter one being obtained from the
former by means of a twist deformation 
\cite{drinfeld,Oeckl:2000eg,Aschieri:2005zs,Aschieri:2005yw}, 
satisfying counit and cocycle condition \cite{majid1}.
The full Hopf algebra description of the theory includes an algebraic as well
 as a coalgebraic part. However, for theories on a commutative spacetime,
the coalgebraic aspects of the symmetry are trivial and basically already
 contained within the algebraic aspects of the symmetry
 transformations. It is for this reason that it suffices to
 describe symmetries only by specifying their Lie algebra structure. On the contrary, for
 theories in noncommutative spacetime, the coalgebraic aspects of
 the symmetry are generally not trivial. So a complete characterization of
 the symmetry requires a description in terms of Hopf algebra.
The Hopf algebra is a quantum symmetry where the issue of 
conserved charges and currents is still subject of research, 
and the Noether theorem generally is still not established. 
However, to obtain a sensible physical picture
out of the Hopf algebra prescription of our theory we will need to do certain  
compromises, most important towards the Noether theorem and momentum conservation. 
We shall elaborate on this issue in the next subsections. Before doing that, the notion of the star product
 still remains to be introduced.

%%%%%%%%%%%%%%%%%%%%%%

\subsection{Coalgebra and star product}

The star product is introduced in the following way:
first let us introduce the unit element $1 \in {\mathcal{A}}$ and define the action of
Poincar\'{e} generators $\p_\mu$ and $M_{\mu \nu}$ on $1$ as
\begin{equation} 
\label{actionde} 
\p_{\mu} \triangleright 1 = 0, \qquad  M_{\mu\nu} \triangleright 1 = 0\,,
\end{equation}
where ${\mathcal{A}}$ is understood as a module for the enveloping algebra
${\mathcal{U}}({\mathfrak{so}}(3,1))~$ of  Poincar\'{e} algebra. In
 other words,
${\mathcal{A}}$ is considered to be a ${\mathcal{U}}({\mathfrak{so}}(3,1))$-module.
Let us consider two noncommutative functions $\hat{\phi}(\x),\hat{\psi}(\x)\in {\hat{\mathcal{A}}}_{\kappa} $. 
Via the vector space isomorphism, these functions are associated to $\phi(x),\psi(x) \in {\mathcal A}$, where
$\phi(x)= \hat{\phi}(\x)\triangleright 1, \psi(x) = \hat{\psi}(\x) \triangleright 1$. 
Then the star product is defined by
\begin{eqnarray} 
\label{starproductdefinition}
\phi (x) \star \psi(x)  ~& :=& ~ \hat{\phi}(\x) \hat{\psi}(\x) \triangleright 1 \nonumber \\
  & = & ~ \hat{\phi}(\x) \triangleright (\hat{\psi}(\x)\triangleright 1)
  ~ = ~ \hat{\phi}(\x)\triangleright \psi(x)\,,
\end{eqnarray}
where it is understood that ${\x}$ is given either by (\ref{2.10}) or
by its hermitian variant (\ref{2.10d}).
In this situation, ${\mathcal{A}}$ considered as a vector space together with 
the star product (\ref{starproductdefinition})
constitutes a noncommutative algebra, which we denote by ${\mathcal{A}}_{\kappa}$.
Unlike the algebra ${\mathcal{A}},$ the algebra ${\mathcal{A}}_{\kappa}$
is isomorphic to the enveloping algebra ${\hat{\mathcal{A}}}_{\kappa}$ 
of NC coordinates. Note that the commutator (\ref{2.1}) can be written in terms of 
ordinary coordinates and $\star$-product (\ref{starproductdefinition}) as
\begin{eqnarray} 
\label{starcommutator}
[x_\mu,x_\nu]_{\star}=x_\mu \star x_\nu - x_\nu \star x_\mu = i(a_\mu x_\nu - a_\nu x_\mu)\,.
\end{eqnarray}

In the familiar context of theories on commutative spacetime, we
  describe a symmetry as a transformation of the coordinates that
  leaves the action of the theory invariant. We keep this notion also
  in case of noncommutative spacetime.
  The symmetry
 underlying the $\kappa$-deformed Minkowski space is the deformed Poincar\'{e} symmetry which can most
 conveniently be described in terms of Hopf algebras. 
 As can be seen in relations (\ref{2.2}), (\ref{2.4}) and (\ref{2.5}),
 the algebraic sector of this deformed symmetry is the same as that of the undeformed 
Poincar\'{e} algebra. However, the action of the Poincar\'{e} generators
 on the deformed Minkowski space is modified in such a way, that the whole deformation
 is contained in the coalgebraic sector. This means that the Leibniz
 rules, which describe the action of the generators $M_{\mu\nu}$ and $p_{\mu}$ 
 on a product of fields,
will no more have the standard form, but instead will be deformed
and will depend on the presentation of $\Phi_{\mu\nu}$. The Hopf algebra structure
 describes the properties of the generators of a deformed
  Poincar\'{e} symmetry. Its algebraic
  sector is determined by the relations (\ref{2.2}), (\ref{2.4}) and (\ref{2.5}).
On the other hand, the coalgebraic sector is determined by the coproducts 
for translation ($p_{\mu}= -i\p_{\mu}$), rotation and boost generators ($M_{\mu\nu}$)
\cite{Meljanac:2007xb,KresicJuric:2007nh},
\begin{eqnarray} 
\label{coproductmomentum}
\triangle \p_{\mu} &=& \p_{\mu}\otimes Z^{-1}+\mathbf{1}\otimes
\p_{\mu}+ia_{\mu} (\p_{\lambda} Z)\otimes
\p^{\lambda}-\frac{ia_{\mu}}{2} \square\, Z\otimes ia\p\,, \\
\triangle M_{\mu\nu} &=& M_{\mu\nu}\otimes
\mathbf{1}+\mathbf{1}\otimes M_{\mu\nu} \nonumber \\
&+& ia_{\mu}\left(\p^{\lambda}-\frac{ia^{\lambda}}{2}\square\right)\,
Z\otimes
M_{\lambda\nu}-ia_{\nu}\left(\p^{\lambda}-\frac{ia^{\lambda}}{2}\square\right)\,
Z\otimes M_{\lambda\mu} \,.
\label{coproductangmomentum}
\end{eqnarray}
In the above expressions, $Z$ is the shift operator, determined by (\ref{16}). 
It has a simple expression for the coproduct:
\begin{equation}
\label{ZZ}
\triangle Z = Z\otimes Z\,.
\end{equation}
The operator $\square$ is the deformed d'Alembert operator 
\cite{Dimitrijevic:2003wv,Meljanac:2006ui,KresicJuric:2007nh},
\begin{equation} 
\label{3.7}
  \square = \frac{2}{a^2}(1- \sqrt{1- a^2 \p^2})\,, 
\end{equation}
which in the limit $a \rightarrow 0$ acquires the
standard form, $\square \rightarrow \p^2$.

The Hopf algebra in question also has  well defined counits and
antipodes. The antipodes for  the generators of the $\kappa$-Poincar\'{e} Hopf algebra 
are given by
\begin{equation} 
\label{3.8}
 S(\p_{\mu}) = \left( -\p_{\mu} + i a_{\mu} {\p}^2 +
 \frac{1}{2}a_{\mu}(a\p)\square \right) Z\,,
\end{equation}
\begin{equation} 
\label{3.8b}
 S(M_{\mu \nu}) = -M_{\mu \nu}
   + ia_{\mu} \left ( \p_{\alpha} - \frac{ia_{\alpha}}{2} \square
   \right ) M_{\alpha \nu}
  - ia_{\nu} \left ( \p_{\alpha} - \frac{ia_{\alpha}}{2} \square 
   \right ) M_{\alpha \mu}\,,
\end{equation}
 while the counits remain trivial.
In the above relations, the operator $Z$ is given by
\begin{equation} 
\label{3.8a}
   Z \equiv \frac{1}{Z^{-1}} = \frac{1}{ -ia\p + \sqrt{1 - a^2 {\p}^{2}}}\,,    
\end{equation}
in accordance with (\ref{16}),
and $\square$ is given in (\ref{3.7}).

Since we are interested in the perturbative expansion of the field
theoretic action,
for later convenience we provide a series expansion up to second order in the deformation parameter $a$ for the expression $\triangle \p_{\mu}$:
\begin{eqnarray} 
  \triangle \p_{\mu} &=&  \p_{\mu}\otimes \mathbf{1} +\mathbf{1}\otimes
 \p_{\mu} 
  - i\p_{\mu} \otimes a\p +
 ia_{\mu} \p_{\alpha} \otimes \p^{\alpha}  
 \nonumber \\
 & -& \frac{1}{2}a^2\p_{\mu}\otimes \p^2 
  - a_{\mu} (a\p)\p_{\alpha} \otimes \p^{\alpha}
   +\frac{1}{2} a_{\mu} \p^2 \otimes a\p
    + {\mathcal{O}}(a^3)\,. 
  \label{Dd}
\end{eqnarray} 

Once we have the coproduct (\ref{coproductmomentum}), we can straightforwardly 
 construct a star product between two arbitrary
fields $f$ and $g$ of commuting coordinates \cite{Meljanac:2006ui,Meljanac:2007xb}.
 For the noncommutative spacetime (\ref{2.1}), the star product has the following form:
\begin{equation} 
\label{d88}
(f \; \star \; g)(x)  =   \lim_{\substack{u \rightarrow x  \\ y \rightarrow x }}
 {\cal M} \left ( e^{x^{\mu} ( \triangle - {\triangle}_{0}) {\partial}_{\mu} }
    f(u) \otimes g(y) \right )\,, 
\end{equation}
where $ {\triangle}_{0}{\partial}_{\mu} =
 {\partial}_{\mu} \otimes 1 + 1 \otimes {\partial}_{\mu}$,
$ \: \triangle ({\partial}_{\mu})$ is given by (\ref{coproductmomentum}),
and $ \; {\cal M} \; $ is the multiplication map in the undeformed Hopf algebra,
 namely $ \; {\cal M} (f(x) \otimes g(x)) = f(x) g(x) ~$ \cite{majid1}.
From (\ref{d88}) we see that the star product only depends on the
 coproduct for translation generators. Its form 
 does not depend on the presentation for $\Phi_{\mu \nu}$. However, since the coproducts depend on $\Phi_{\mu \nu}$
 so does the star product according to (\ref{d88}), 
 in an implicit form. At this point we emphasize here once
 again that in this paper we are doing analysis based on the specific
 realization (\ref{2.10}) and its hermitian
 variant (\ref{2.10d}), given in the next section.
 The coproducts (\ref{coproductmomentum}) and (\ref{coproductangmomentum})
are written for and correspond to this particular type of realization. One can check
 that the star product (\ref{d88}) with the coproduct
 (\ref{coproductmomentum}) is associative.
 
\section{Modified $\kappa$-deformed scalar field action}
\subsection{Nonhermitian realization of the NC $\phi^4$ action}

In this subsection, we construct an interacting scalar field theory on
 noncommutative spacetime whose short distance geometry is governed by
 the $\kappa$-deformed symplectic structure (\ref{2.1}). In
 particular, we are interested in a field theoretic action describing
 the dynamics of a massive scalar field, i.e. generalizing the celebrated
 Grosse-Wulkenhaar action in dimension $n$,
\begin{eqnarray} 
\label{actionstar}
S_n[\phi]&=&\int d^n x~{\mathcal{L}}(\phi,\p_{\mu}\phi,\p_{\mu}\p_{\nu}\phi)\nonumber\\
 & = & \frac{1}{2} \int d^n x ~ (\p_{\mu}\phi) \star (\p^{\mu}\phi) + 
  \frac{m^2}{2}\int d^n x ~ \phi \star \phi  
   + \frac{\xi^2}{2} \int d^n x ~ x_{\mu}\phi \star x^{\mu} \phi 
   \nonumber \\    
 & + & \frac{\lambda}{4!} 
 \int d^n x ~ \phi \star \phi \star \phi \star \phi\,.
\end{eqnarray}
Note that the $\phi^4$ interaction term is accompanied by an additional harmonic term of
the Grosse-Wulkenhaar type, along with the standard kinetic and mass terms.
Due to the very definition of
the $\star$-product (\ref{d88}), all terms in the action
(\ref{actionstar}) will be  $\kappa$-deformed. In the case of the nonhermitian
realization (\ref{2.10}), the scalar field $\phi$ is, up to first
order in the deformation parameter $a,$ governed by the action
\begin{eqnarray} 
  S_n[\phi] & = & \frac{1}{2}\int d^n x ~ \Big[(\p_{\mu}\phi) (\p^{\mu}\phi) +
 m^2 \phi^{2} + \xi^2 x^{2}\phi^2 + \frac{\lambda}{12} \phi^{4}\Big] 
 \nonumber \\
 & + & \frac{1}{2} a_{\mu}\int d^n x ~ 
 \Big\{ (n-1)  \xi^2   x^{\mu} \phi^2  
 \nonumber \\
 & + & ({\eta}^{\alpha\beta}x^{\mu}-{\eta}^{\mu\beta}x^{\alpha})
 \nonumber \\
 & \times &  
 \Big[\phi x_{\alpha}(\p_{\beta}\phi) 
 + (m^2+\xi^2x^2+\frac{\lambda}{3}\phi^2)(\p_{\alpha}\phi)(\p_{\beta}\phi)+
 (\p_{\alpha}\p_{\gamma}\phi)(\p_{\beta}\p^{\gamma}\phi)\Big]\Big\}\,.
 \label{actionscalar}
\end{eqnarray}

However, in this case the scalar field cannot be properly defined and it
is not clear at all if it is complex or real. 
Next, besides that various terms from (\ref{actionstar}) 
attain an explicit $x$-dependence after the expansion in $a$, 
it is easy to note the lack of a typical property of the Moyal $\star$-product, namely 
$\int f\star g = \int f \cdot g$. 
This makes the above nonhermitian models uncontrollable, thus 
less attractive, since it would be very difficult to extract some physical meaning. Generally, to have a scalar field
defined in the right way and to be able to say whether it is real or
complex, it is necessary to introduce an involution operation $\dagger$ 
corresponding to the adjoint or hermitian
conjugation operation $\ddagger$, which in turn requires to consider and work
with a hermitian, instead of the nonhermitian realization.
The problems just described can thus be properly resolved
only after we replace the nonhermitian realization (\ref{2.10}) with a
hermitian one and introduce the notion
of an antipode in order to properly define an adjoint or hermitian
conjugation operation $\ddagger$.
In order to proceed, we introduce the scalar product $(\cdot,\cdot)$ on the algebra
${\mathcal{A}}_{\kappa}$ as
\footnote{To avoid further confusion, we discuss the issue of different
notation for different operations and their properties; that is, 
the usual scalar product $(f,g)=\int d^n x f^*\cdot g$ induces the usual 
hermitian conjugation operation $\dagger$.
Here $*$ represents the usual complex conjugation. The 
scalar product $(\psi,\phi)_{\kappa}=\int d^n x~{\psi}^{\dagger}~{\star}_h~\phi$ 
on the algebra ${\mathcal{A}}_{\kappa}$, where $\dagger$ is the corresponding involution 
(generalized complex conjugation), 
induces the required hermitian conjugation operation $\ddagger$.  
Note that we denoted two different notions by the same symbol, 
that is $\dagger$ denotes the ordinary hermitian conjugation operation 
$\partial^\dagger=-\partial$ and the involution used to define the scalar product
in ${\mathcal{A}}_{\kappa}$. From the context, it should be clear which one is meant; namely when $\dagger$ acts on a function, it is the involution, and
when it is applied to an operator it is the ordinary hermitian conjugation.
It is understood that $\partial^\ddagger =(S(\partial))^\dagger=-\partial$,  
$M^{\ddagger}_{\mu\nu}=(S(M_{\mu\nu}))^\dagger$, or compactly written
$A^{\ddagger}=(S(A))^\dagger$, for any generator $A$ of the 
$\kappa$-deformed algebra. It has to be stressed that $\dagger$, 
in a previous sentence, means involution with respect 
to the deformed scalar product (\ref{4.21}). Some properties of above operations, 
generalized trace property and behavior with respect to integration by part  
are given in \cite{Meljanac:2010ps}.}
\begin{align} 
\label{4.21}
 (\psi, \phi)_{\kappa} = \int d^n x ~ {\psi }^{\dagger} ~ {\star}_h ~ \phi\,, 
\end{align}
where $\dagger$ is the involution 
with respect to this new scalar product, defined in terms of the star product
$\star_h$, whose explicit form will be provided later.
Assuming that functions in the algebra ${\mathcal{A}}_{\kappa}$
have the Fourier expansion
\begin{equation} 
\label{4.19}
 \phi (x) = \int d^n p ~ \tilde{\phi} (p) ~ e^{ipx},
\end{equation}
the action of the operation $\dagger$
on plane waves is specified as follows
\begin{equation} 
\label{4.18}
 {\left(e^{ipx} \right)}^{\dagger} = e^{iS(p)x}\,.
\end{equation}
Here $S(p) = -iS(\p)$ is the antipode (\ref{3.8}) and $\dagger$ is the 
involution defined on ${\mathcal{A}}_{\kappa}$.
Note that the reality of the field $\phi$ can be defined in a more than one way, 
depending on the conjugation operation we might choose: $\phi^\dagger=\phi$
or $\phi^*=\phi$. When we use the term "real field", we have the first case in mind.
For further details see reference \cite{Meljanac:2010ps}.

In case of Moyal-deformation, we have $(e^{ipx})^{\dagger} = e^{-ipx}$ as in the commutative case instead of (\ref{4.18}). In order to clarify particle-antiparticle
plane waves, one needs to modify the interpretation of (\ref{4.18}) accordingly.
However, this is not trivial for $a\not=0$. 
In our approach the algebraic sector of the Poincare algebra, 
i.e. commutation relations (\ref{2.2}) are undeformed,
thus the corresponding Casimirs are also undeformed,
and the dispersion relation remains the same, $p^2=m^2$.
From the antipodes (\ref{3.8}) 
for the translation generators of the $\kappa$-Poincar\'{e} Hopf algebra, it 
follows that $S^2(p)=p^2=m^2$, in agreement with previous conclusions regarding the
dispersion \cite{Meljanac:2010ps}.

\subsection{Hermitian realization of the NC $\phi^4$ action}

In order to obtain the physical meaning of the NC $\phi^4$ field theory, we have to 
introduce a complex scalar field $\phi$ together with the corresponding notion of
hermitian conjugation, as explained above and will proceed further with the construction
of a hermitian theory. 

In order to obtain a hermitian action, we are necessarily forced to work with 
a hermitian realization represented by operators ${\x}_{\mu}^{h}$, with the property
${({\x}_{\mu}^{h})}^{\dagger} = {\x}_{\mu}^{h}.$ These  fully
hermitian operators $ {\x}_{\mu}^{h}$ can be constructed from the operators (\ref{2.10}) as
\begin{equation} 
{\x}_{\mu}^{h}=\frac{1}{2}(\x_{\mu} + \x_{\mu}^{\dagger})={({\x}_{\mu}^{h})}^{\dagger}\,,
\label{2.10H}
\end{equation}
which results in ($\dagger$ here means usual hermitian conjugation, 
$x_{\mu}^\dagger=x_{\mu},\;\partial_{\mu}^\dagger=-\partial_{\mu}$)
\begin{equation} 
\x_{\mu}^{h} = 
 x_{\mu} \sqrt{1+ a^2 p^2 } - iM_{\mu \nu} a^{\nu} - 
 i \frac{a^2}{2} \frac{1}{\sqrt{1+ a^2 p^2 }} p_{\mu}\,.
\label{2.10d}
\end{equation}
Changing the realization of the coordinates from (\ref{2.10}) to (\ref{2.10d}) 
%in accordance with (\ref{starproductdefinition}), 
modifies the form of the star product, and we obtain a the new star product denoted as ${\star}_h$:
 \begin{eqnarray} 
 \label{starproductdefinitionher}
 \phi(x){\star}_h\psi(x)~& =&~\hat{\phi}({\x}^{h})\hat{\psi}({\x}^{h})\triangleright 1 
 \nonumber \\
  & = & ~ \hat{\phi}({\x}^{h}) \triangleright (\hat{\psi}({\x}^{h}) \triangleright 1)
  ~ = ~ \hat{\phi}({\x}^{h})\triangleright \psi(x)\,,
\end{eqnarray}
with ${\x}^{h}$ being given by (\ref{2.10d}).  
Thus, we are forced to replace the star product (\ref{d88}) with
a new one of the following form \cite{Meljanac:2010ps}:
\begin{equation} 
\label{newstarproduct}
(f \; {\star}_h \; g)(x)  =   \lim_{\substack{u \rightarrow x  \\ y \rightarrow x }}
 {\cal M} \left ( e^{x^{\mu} ( \triangle - {\triangle}_{0}) {\partial}_{\mu} }
  \sqrt[4]{\frac{1- a^2
   \triangle ( {\p}^2)}{(1- a^2 ~ {\p}^2 \otimes 1) ~ (1-
   a^2 ~ 1 \otimes {\p}^2 )}} ~
  f(u) \otimes g(y) \right )\,, 
\end{equation}
where it is understood that the coproduct $\triangle ({\p}_{\mu})$,
 Eq.(\ref{coproductmomentum}), is a homomorphism, i.e.
 $\triangle ({\p}^2) = \triangle ({\p}_{\mu}) \triangle ({\p}^{\mu})$.
In this way, the nonhermitian version of the star product (\ref{d88}) is
replaced by the above hermitian one, (\ref{newstarproduct}).

The star product ${\star}_h$ is associative. However, in contrast to the star product (\ref{d88}), 
it has the following desirable property \cite{Meljanac:2010ps}:
\begin{eqnarray} 
\int d^n x \;\phi^{\dagger} {\star}_h \psi = \int d^n x \;\phi^{*} \cdot \psi\,,
\label{property}
\end{eqnarray}
where the asterisk $*$ denotes the usual complex conjugation.

The above results -- the new ${\star}_h$-product (\ref{newstarproduct}), and 
the identity (\ref{property}) -- encode a very nice and important property:
{\it the integral measure problems are avoided}.
That means that by usage of the ${\star}_h$-product,
corresponding to the hermitian realization of the $\kappa$-Minkowski spacetime,
the measure function is naturally absorbed 
within the new ${\star}_h$-product.  
Therefore, the action (\ref{actionstar}) has to be replaced by
\begin{eqnarray} 
\label{actionnew}
 S_n[\phi] & = & \int d^n x ~ {\mathcal{L}}(\phi,\p_{\mu}\phi,\p_{\mu}\p_{\nu}\phi)
 \nonumber \\
 & = & \int d^n x ~ (\p_{\mu}\phi)^{\dagger} {\star}_h (\p^{\mu}\phi) + 
  m^2 \int d^n x ~ \phi^{\dagger} {\star}_h \phi  
   + \xi^2 \int d^n x ~ x_{\mu}\phi^{\dagger} {\star}_h x^{\mu} \phi  
\nonumber \\    
& + &\frac{\lambda}{4}  
\int d^n x~\frac{1}{2}(\phi^{\dagger}{\star}_h\phi^{\dagger}{\star}_h \phi {\star}_h \phi 
    +\phi^{\dagger} {\star}_h \phi {\star}_h \phi^{\dagger} {\star}_h \phi)\,.
\end{eqnarray}
In the above action, the interaction $\phi^4$ term should in fact
incorporate six terms corresponding to all possible permutations of
the fields $\phi$ and $\phi^{\dagger}.$ However, due to the integral property
(\ref{property}), these six
different permutations can be reduced to only two mutually nonequivalent terms, namely
 $\phi^{\dagger} {\star}_h \phi^{\dagger} {\star}_h \phi {\star}_h \phi$ and
 $\phi^{\dagger} {\star}_h \phi {\star}_h \phi^{\dagger} {\star}_h \phi$.

When expanded up to first order in the deformation parameter $a,$
the action (\ref{actionnew}) after some rearrangements including
integration by parts, receives the following form
\begin{eqnarray} 
  S_n[\phi] 
 & = & \int d^n x ~ \Big[(\p_{\mu}\phi^{*}) (\p^{\mu}\phi) +
 (m^2 + \xi^2 x^{2})\phi^{*}\phi + \frac{\lambda}{4}{(\phi^{*}\phi)}^2\Big] 
 \nonumber \\
 & + & 
 i\frac{\lambda}{4}\int d^n x \bigg[ a_{\mu} ~ 
 x^{\mu} \Big({\phi^{*}}^2(\p_{\nu} \phi) {\p}^{\nu} \phi -  
 \phi^{2}(\p_{\nu} {\phi}^{*}) {\p}^{\nu} {\phi}^{*}\Big)  
\nonumber \\
 & + & 
 a_{\nu} ~ x^{\mu} \Big(\phi^{2}(\p_{\mu} \phi^{*}) {\p}^{\nu} \phi^{*} -  
 {\phi^{*}}^2(\p_{\mu} {\phi}) {\p}^{\nu} {\phi}\Big)  
  \nonumber \\
  &  + &       
    \frac{1}{2}  a_{\nu} x^{\mu}~ \phi^{*}\phi \Big( 
    (\p_{\mu} {\phi}^{*}) {\p}^{\nu} \phi
     -(\p_{\mu} \phi) {\p}^{\nu} {\phi}^{*}\Big)  \bigg ]
    + {\mathcal{O}}(a^2)\,.
 \label{actioncomlex}
\end{eqnarray}
Note that the oscillator term proportional to $\xi^2$ attain no correction 
in the deformation parameter $a$. 
These two features of our model separate completely in the above action.

At the end of this subsection note that the underlying Hopf algebra  
is a twisted symmetry algebra, where existence/conservation of charges and currents
are still subject of research. 
However the action (\ref{actioncomlex}), obtained by expansion of (\ref{actionnew}) 
up to first order in the deformation parameter $a$,
is invariant under the following internal symmetry transformations:
\begin{equation}
\left(
    \begin{array}{c}
        \phi \\
       \phi^{\ast}
    \end{array}
\right)
\rightarrow
%\end{equation}
%\begin{equation}
\begin{pmatrix}
    e^{i\chi} &0 \\
   0 & e^{-i\chi}
\end{pmatrix}
\left(
    \begin{array}{c}
        \phi \\
       \phi^{\ast}
    \end{array}
\right)\,,
\label{intsym}
\end{equation}
thus the corresponding Noether current should be conserved.

\subsection{Equations of motion and Noether currents of internal symmetry}

We proceed further by evaluating the equations of motion for the fields
$\phi$ and ${\phi}^{\ast}$: 
\begin{eqnarray} 
\label{eom2}
 {\p}_{\mu} {\p}^{\mu} \phi - 
 (m^2 + {\xi}^2 x^2) \phi  & = & \frac{\lambda}{6}  
 \Big[{\phi}^{\ast} {\phi}^2 + 
  ia_{\mu} x^{\mu}\Big( {\phi}^2 
 {\p}_{\nu} {\p}^{\nu} {\phi}^{\ast} + 
   {\phi}^{\ast}
 ({\p}_{\nu} \phi) {\p}^{\nu} \phi
  + \phi 
 ({\p}_{\nu} {\phi}) {\p}^{\nu} {\phi}^{\ast} \Big)  
 \nonumber\\
 & - & ia^{\mu}x^{\nu} \Big( {\phi}^{\ast} 
 ({\p}_{\nu} \phi) {\p}_{\mu} \phi 
 + \phi 
 ({\p}_{\nu} \phi) {\p}_{\mu} {\phi}^{\ast}
 +  {\phi}^2 
 {\p}_{\mu} {\p}_{\nu} {\phi}^{\ast} 
  +  \phi ({\p}_{\mu} \phi)  {\p}_{\nu} {\phi}^{\ast}\Big)  
 \nonumber \\    
 & + & \frac{i}{4} (1-n) a^{\mu} {\phi}^{\ast} \phi 
 {\p}_{\mu} \phi + \frac{i}{2} (1-n) a^{\mu} {\phi}^2 
 {\p}_{\mu} {\phi}^{\ast} \Big]\,,
\end{eqnarray}
where $\phi = 0$ is the trivial solution of the above equation, as it should be.
The equation of motion for ${\phi}^{\ast}$ can be obtained from (\ref{eom2}) by complex conjugation.
 
Next, we present Noether currents derived from 
the Lagrangian densities (\ref{actioncomlex}):
\begin{eqnarray}
j^{\mu}(x)&=&i\frac{\delta{\cal L}}{\delta(\p_\mu \phi)}\phi -
i\frac{\delta{\cal L}}{\delta(\p_\mu {\phi}^{\ast})}{\phi}^{\ast}\,,
\label{current}\\
\frac{\delta{\cal L}}{\delta(\p_\mu \phi)}
&=&
\frac{1}{2}\p^{\mu}{\phi}^{\ast}+ 
\frac{i\lambda}{4} 
\Big[{{\phi}^{\ast}}^2 (2a_{\nu}x^{\nu} \p^{\mu}-a^{\nu}x^{\mu}\p_{\nu}
-a^{\mu}x^{\nu}\p_{\nu})\phi +
\frac{1}{2} {\phi}^{\ast}\phi (a^{\mu}x^{\nu}\p_{\nu}-
a^{\nu}x^{\mu}\p_{\nu}){\phi}^{\ast} \Big]\,,
\nonumber\\
\frac{\delta{\cal L}}{\delta(\p_\mu {\phi}^{\ast})}
&=&
\frac{1}{2}\p^{\mu}{\phi} - 
\frac{i\lambda}{4} 
\Big[{\phi}^2 (2a_{\nu}x^{\nu} \p^{\mu}-a^{\nu}x^{\mu}\p_{\nu}
-a^{\mu}x^{\nu}\p_{\nu})\phi^{\ast} +
\frac{1}{2} {\phi}^{\ast}\phi (a^{\mu}x^{\nu}\p_{\nu}-
a^{\nu}x^{\mu}\p_{\nu}){\phi} \Big]\,.
\nonumber
\end{eqnarray}
The above current (\ref{current}) is conserved; that is 
 \begin{equation}
 \p_{\mu}j^{\mu}(x)=0\,, 
 \label{cc}
 \end{equation}
as it should be due to the invariance of the Lagrangian (\ref{actioncomlex}) 
under the internal symmetry 
transformations (\ref{intsym}).
This can be shown by 
straightforward computations from (\ref{actionnew})-(\ref{cc}).

\section{Quantum properties of the model}

\subsection{Feynman rules}

Even though the S-matrix LSZ formalism, including the Wick theorem, is not quite clearly 
defined on $\kappa$-Minkowski noncommutative spacetime, we continue bona fide towards the 
research of the quantum properties of the model defined by the action (\ref{actioncomlex}).
To do that, we first derive the relevant Feynman rules and than compute the tadpole diagram
contributions to the 2-point Green's function of our model.

Due to the $\kappa$-deformation of our theory,  
the statistics of particles is twisted, 
so that we are generally no more dealing with pure bosons.
We are in fact dealing with {\it something} whose statistics is governed by 
the statistics flip operator
\cite{Balachandran:2005eb,Balachandran:2006pi,Balachandran:2010xk,Balachandran:2010wq} 
and the quasitriangular structure
(universal R-matrix) on the corresponding quantum group 
\cite{majid1,vgdrinfeld,faddeev,Rmatrixmajid}. 
It would be interesting to investigate  
these mutual relations more thoroughly,
but at  the surface level, we can argue that it is possible 
to pick up the basic characteristics of the twisted statistics by using 
the nonabelian momentum addition law 
\cite{Kosinski:1999dw,AmelinoCamelia:2001fd,Daszkiewicz:2004xy,Freidel:2006gc,
KowalskiGlikman:2009zu,KowalskiGlikman:2004qa,
AmelinoCamelia:2001me,KowalskiGlikman:2002ft,Lukierski:2002wf}. 
It can be seen that the accordingly induced deformation of the $\delta$-function 
(arising from the implementation of the nonabelian momentum
addition/subtraction rule) yields the usual $\delta$-function 
multiplied by a certain statistical factor,
which could have its origin in $\kappa$-modified statistics.
When we speak about deformed statistics, we have a less rigid
notion of statistics in mind than applied to the symmetry properties of states,
where multiparticle change of 4-momenta may change the state's symmetry properties.

In order to obtain the Feynman rules in momentum space, 
we are suggesting to use the following line of reasoning,
which we shall from now on call {\it hybrid approach}. \\
(A) We use the methods of standard QFT and treat the modifications 
in action (\ref{actioncomlex}) as a perturbation. 
In doing this, we obtain propagators and Feynman rules for vertices. \\
(B) We know that 
the statistics of particles is twisted and that it has to 
be implemented into the formalism.
Thus we require that the ordinary addition/subtraction rule induces
the addition rule for twisted statistics on $\kappa$-Minkowski spacetime.
In momentum space, this means 
\begin{equation}
\sum_i k^{\mu}_i\;\; \rightarrow \;\;
\sum_{\oplus i} k^{\mu}_i  \;\;\;\;\,\&\,\;\;\;\;
\sum_i k^{\mu}_i - \sum_j p^{\mu}_j\;\; \rightarrow \;\;
\sum_{\oplus i} k^{\mu}_i \ominus \sum_{\oplus j} p^{\mu}_j \,,
\label{oplus+}
\end{equation}
where induced $direct$ addition/subtraction rules are going to be
defined in Subsection 4.1.2, for the simplest cases of two and four momenta.   
%Note that the associativity for the direct sum $\oplus$ is satisfied due to the associativity of the star product (\ref{newstarproduct}).
We proceed in two steps:\\
(1) Following the arguments above, we implement 
the induced conservation law within the $\delta$-functions in 
the Feynman rule, and\\
(2) whenever needed, we use the modified/deformed conservation law along the course
of evaluation of the Feynman diagrams.

\subsubsection{Feynman rules (A): standard momentum addition law}

From now on we will restrict ourselves to the Euclidean metric.
Note that for the transition from Minkowski to Euclidean signature,
we are using the following rules: 
$a^M=(a^M_0,a^M_i) \longrightarrow a^E=(a^E_i, a^E_n)$,
where $a^E_n=ia_0^M$, and similary for any $n$-vector. Thus scalar product
is defined as $a^E k^E=a^E_{\mu}k^E_{\mu}=a^E_i k^E_i + a^E_n k^E_n
=-a^M_0 k^M_0 + a^M_i k^M_i= a^M k^M$. 
In subsequent consideration we drop the $M$ and $E$ superscripts, 
but it is understood that we work with Euclidean ($E$) quantities.
As the quadratic part of the action (\ref{actioncomlex})
in $\kappa$-space is modified by the harmonic oscillator term,
the propagator in momentum space is also going to be modified,
\begin{eqnarray}
 \tilde \Gamma^{\xi} (k_1,k_2) &=& 
 -i\Bigg(\frac{\delta^2 S[\tilde \phi]}{\delta \tilde \phi^*(k_1) 
 \delta \tilde \phi(k_2)}\Bigg)_{\tilde \phi = \tilde \phi^*=0}\\ 
 \nonumber
 &=& -i(2\pi)^{n}\bigg[(k_{1\mu} k_{2\mu} + m^2)
 +\xi^2\partial^2_{k_1}\bigg]\delta^{(n)}(k_1-k_2)\,,
 \label{prop}
\end{eqnarray}
with all fields $\phi,\phi^*$ having the same nonzero mass. 
The free propagator is than symbolically written as
\begin{equation}
{\tilde G}^{\xi}(k_1,k_2) = \frac{(2\pi)^n}{\tilde \Gamma^{\xi}(k_1,k_2)}
=i\Bigg((k_{1\mu} k_{2\mu}+m^2+\xi^2\partial^2_{k_1})\,\delta^{(n)}(k_1-k_2)\Bigg)^{-1}\,.
\label{propagator}
\end{equation}
Since the harmonic oscillator term in the action (\ref{actioncomlex}) breaks translational invariance, it is clear from above 
that translational invariance breaking produces some kind of modification of the mass. 
The parameter $\xi^2$ has dimension $dim[\xi]=length^{-2}$ and 
specifies the magnitude of the breaking of translational invariance. 

If one includes a harmonic oscillator term, the propagator is given by 
the so-called Mehler kernel which depends on
different incoming and outgoing momenta since the propagator does not respect
momentum conservation \cite{Blaschke:2009aw}. 
So for full computation, one needs to take into account the
Mehler kernel from the beginning. 
However, in the spirit of our {\it hybrid approach} and under
the assumption of small perturbation due to the harmonic term, 
we approximate the full propagator by
\begin{equation}
G^{\xi}\equiv G^{\xi}(k_1,k_2) \simeq\frac{i}{k_1^2+m^2}
 \Bigg(1-\frac{\xi^2}{(k_1^2+m^2)}\partial^2_{k_1}\Bigg)\delta^{(n)}(k_1-k_2)\,.
 \label{propagator1}
\end{equation}
This is going to be used in the computation of the 2-point Green's function.
We believe that the approximative expression (\ref{propagator1}) is good enough
to help us to indicate the influence of 
the $\xi^2$ term on the one-loop quantum corrections.
This is going to be presented in Subsection 4.2.
Of course the full computation of quantum corrections including 
the Mehler kernel is out of
scope for this paper, but it is certainly going to be performed in the future.

If $a$ is of the order of the Planck length, $\xi$, 
despite being small, carries contributions to Green's functions
that are  still larger than the terms linear in $a$.

The vertex function, which in momentum space is given by
\begin{equation}
\tilde\Gamma (k_1,k_2,k_3,k_4;a)=i\frac{\delta^4 S[\tilde \phi]}{\delta\tilde\phi(k_1) 
\delta\tilde\phi(k_2)\delta\tilde\phi^*(k_3) \delta \tilde \phi^*(k_4)} \,,
\label{vertex}
\end{equation}
is modified too. 
It is illustrated in Fig.~\ref{fig:vertex} and amounts to the following expression:
\begin{figure}
%\FIGURE{figure}
\begin{center}
\includegraphics[width=40mm]{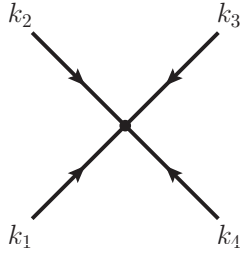}
\end{center}
\caption{Scalar 4-field vertex}
\label{fig:vertex}
\end{figure}

\begin{eqnarray}
\nonumber
 \tilde\Gamma(k_1,k_2,k_3,k_4;a) &= &i(2\pi)^n\frac{\lambda}{2}\,a_{\nu} 
\Bigg[\frac{a_\nu}{a^2}+ 
\frac{1}{4}\bigg(k_{4\mu} k_{3\nu}+k_{3\mu}k_{4\nu}-2\delta_{\mu\nu}k_{4\rho}k_{3\rho}\\
\nonumber
&&\hspace{-2.5cm}
+\frac{1}{2}(k_{2\mu}k_{4\nu}-k_{4\mu}k_{2\nu}+k_{2\mu}k_{3\nu}-k_{3\mu}k_{2\nu})\bigg)
\partial_\mu^{k_1}\\
\nonumber
&& \hspace{-3cm}
+ \frac 14 \bigg( k_{4\mu} k_{3 \nu} + k_{3\mu} k_{4\nu} 
- 2 \delta_{\mu\nu} k_{4\rho} k_{3\rho} \\
\nonumber
&& \hspace{-2.5cm}
+ \frac 12 ( k_{1\mu}k_{4\nu} - k_{4\mu}k_{1\nu} + k_{1\mu}k_{3\nu} - k_{3\mu}k_{1\nu} )
\bigg) \partial_\mu^{k_2}\\
%%
%\nonumber
&& \hspace{-3cm}
+ \frac 14 \bigg(
k_{1\mu}k_{2\nu} + k_{2\mu}k_{1\nu} - 2 \delta_{\mu\nu} k_{1\rho}k_{2\rho}
\bigg) ( \partial_\mu^{k_3} + \partial_\mu^{k_4} )
\Bigg] \delta^{(n)} (k_1 + k_2 - k_3 - k_4),
\label{Feynrule}
\end{eqnarray}
where we denote $\partial_\mu^k = \frac{\partial}{\partial k_\mu}$, 
and all four momenta $k_i$ are flowing into the vertex.
Both couplings, $\xi$ and $\lambda$ have to be dimensionally regularized.

\subsubsection{Feynman rules (B): $\kappa$-deformed momentum addition law}

Next, we discuss the notion which anticipates the 
induced momentum conservation law on $\kappa$-space, 
within our {\it hybrid approach}. 
Namely, the $\delta$-function in (\ref{Feynrule})  
comes from the contraction of fields, where 
the momentum conservation should be obeyed in accordance with
the $\kappa$-deformed momentum addition rule 
\cite{Kosinski:1999dw,AmelinoCamelia:2001fd,Freidel:2006gc}.
We have two cases for summation/subtraction of 4-vectors $k^{\mu}$
with respect to the physical situation of four particles and/or quantum fields
propagating in space with respect to an interaction point:\\
(I) all particle momenta flowing into the vertex, as given in Fig. \ref{fig:vertex}
\begin{equation}
k_{1\mu} + k_{2\mu} + k_{3\mu} + k_{4\mu} =0\;\; \to \;\;
k_{1\mu} \oplus k_{2\mu} \oplus k_{3\mu} \oplus k_{4\mu} =0\,,
\label{oplus}
\end{equation}
(II) the process of scattering "2 particle $\to$ 2 particle", where we have 
\begin{equation}
(k_{1\mu} + k_{2\mu}) - (k_{3\mu} + k_{4\mu}) =0\,
\;\; \to \;\;
(k_{1\mu} \oplus k_{2\mu}) \ominus (k_{3\mu} \oplus k_{4\mu}) =0\,.
\label{ominus}
\end{equation}

Having obtained the Feynman rules (\ref{propagator1}) and (\ref{Feynrule}),
we have completed the first stage of our program, that is to deduce
the free propagation and interaction properties of the model by using the standard
quantization.
 
At this point we turn to the second part, which 
includes the effective description of the statistics of  particles
described by the model. As already indicated before, the statistics of
particles is twisted in $\kappa$-space 
\cite{Govindarajan:2009wt,Young:2007ag,Govindarajan:2008qa,Young:2008zg},
with the deformation being encoded within the nonabelian momentum addition rule. It is
known that the rule for addition of momenta is governed by the
coproduct structure of the Hopf algebra in question. In our case,
the relevant Hopf algebra is the $\kappa$-Poincar\'{e} algebra and the
corresponding coalgebra structure is given by (\ref{coproductmomentum}),
 (\ref{coproductangmomentum}), and (\ref{3.8}), (\ref{3.8b}), respectively.
In particular, the coproduct (\ref{coproductmomentum}) for translation
generators determines the required momentum addition rule,  
which in momentum space and up to the first order in
deformation $a$, from (\ref{3.8}) and (\ref{Dd}), yields: 
\begin{eqnarray}
 S(p_{\mu}) &=&-iS(\partial_{\mu})=-p_{\mu}-a_{\mu}p^2+(ap)p_{\mu} 
 +{\mathcal O}(a^2)\,,
\label{4} \\
(p_{\mu} \oplus k_{\mu}) &=& (p+k)_{\mu} + (ak)p_{\mu} - a_{\mu}(p k)
+{\mathcal O}(a^2)\,.
\label{4+}
\end{eqnarray}
The second equation (\ref{4+}) states the nonabelian momentum addition rule, while
$S(p)$ is the antipode with the property 
$p^{\mu} \oplus S(p^{\mu}) = 0$. 
%, that in fact represents the very definition of the antipode. 
Since commutativity in momentum space is not satisfied, 
i.e. $k\oplus p \not= p\oplus k$, a certain {\it ordering} has to be implemented.
However, instead of implementing a possibly {\it complicated unknown ordering},
we shall proceed in the most simple way by taking into account all possible type 
of contributions; for example $k\oplus p\oplus q$, $p\oplus k \oplus q$, etc.
Combining (\ref{4}) and (\ref{4+}), 
we obtain the following momentum subtraction rule:
\begin{eqnarray}
p_{\mu} \ominus k_{\mu}  &\equiv& (p \oplus S(k))_{\mu} = 
 (p-k)_{\mu}(1 - ak) + a_{\mu}(p k - k^2)
 +{\mathcal O}(a^2)\,.
\label{5}
\end{eqnarray}
This enables us to rewrite the energy-momentum conservation which is
supposed to be satisfied at each vertex. Thus, if two external momenta
$k_1$ and $k_2$ flow into the vertex and the other two external
momenta $k_3$ and $k_4$ flow out of the vertex, then, writing in components, we have
the induced momentum conservation law (\ref{ominus}),
which corresponds to our physical situation .
%while computing 4-field tadpole diagram in the next subsection.

In order to obtain the expressions for the $\delta$-functions in 
Feynman rules, we are starting with
\begin{equation}
\delta^{(n)}(p \ominus k) = \sum_i
\Bigg|\det \Bigg(\frac{\partial {(p \ominus k)}_{\mu}}{\partial p_{\nu}}\Bigg)_{p = q_i} 
\Bigg|^{-1}
\delta^{(n)}(p-q_i) \,,
\label{6}\end{equation}
where we have to sum over all zeros $q_i$ for 
the expression in the argument of the $\delta$-function. 
Since there is only one zero at $q_i=k$, with the help of subtraction rule (\ref{5}),
we find the following first order contribution to the above
$\delta^{(n)}$-function
\begin{equation}
\delta^{(n)}(p \ominus k) = \frac{\delta^{(n)}(p - k)}{{(1 - ap)}^{n-1}} = 
   (1 + (n-1)ap + \mathcal{O}(a^2))\delta^{(n)}(p -k) \,.
\label{7}
\end{equation}

It was shown in \cite{Meljanac:2010ps} that the star product $\star_h$ 
(\ref{newstarproduct}) breaks translation invariance
(in the sense of \cite{Kosinski:1999dw}). 
However, this feature does not show up 
until computations are extended to second order in
the deformation parameter $a$.
The important point is that translation invariance is conserved within
the first order deformation. Since we are carrying out our investigation
in exactly this order, we are allowed to invoke the energy momentum conservation
albeit in a modified form, dictated by the modified 
coproduct structure and by the oscillator term.
Relations between Hopf algebra symmetries
and conservation laws are an important subject of investigation. 
This is the issue of generalizing the Noether theorem,
thus the whole subject is still appealing \cite{Agostini:2006nc}.

With the idea of implementing the new physical features that have just been described,
we modify the Feynman rule (\ref{Feynrule}).
Taking into account all possible contributions, with the help of 
(\ref{4+}) and (\ref{5}), and
choosing the following replacement of $\delta$-function from (\ref{Feynrule})
\cite{Kosinski:1999ix,AmelinoCamelia:2001fd,Daszkiewicz:2004xy,Kosinski:2003xx},
\begin{equation}
\delta^{(n)} (k_1 + k_2 - k_3 - k_4) \rightarrow
\delta^{(n)} ((k_1 \oplus k_2) \ominus (k_3 \oplus k_4))
+\delta^{(n)} ((k_1 \oplus k_2) \ominus (k_4 \oplus k_3))\,,
\label{delta}
\end{equation}
we obtain the {\it hybrid} Feynman rule which obeys the 
$\kappa$-deformed momentum addition/subtraction rule
via sum of $\delta$-functions (\ref{delta}):
\begin{eqnarray}
\nonumber
 \tilde\Gamma(k_1,k_2,k_3,k_4;a) &= &i(2\pi)^n\frac{\lambda}{2}\,a_{\nu} 
\Bigg[\frac{a_\nu}{a^2}+ 
\frac{1}{4}\bigg(k_{4\mu} k_{3\nu}+k_{3\mu}k_{4\nu}-2\delta_{\mu\nu}k_{4\rho}k_{3\rho}\\
\nonumber
&&\hspace{-2.5cm}
+\frac{1}{2}(k_{2\mu}k_{4\nu}-k_{4\mu}k_{2\nu}+k_{2\mu}k_{3\nu}-k_{3\mu}k_{2\nu})\bigg)
\partial_\mu^{k_1}\\
\nonumber
&& \hspace{-3cm}
+ \frac 14 \bigg( k_{4\mu} k_{3 \nu} + k_{3\mu} k_{4\nu} 
- 2 \delta_{\mu\nu} k_{4\rho} k_{3\rho} \\
\nonumber
&& \hspace{-2.5cm}
+ \frac 12 ( k_{1\mu}k_{4\nu} - k_{4\mu}k_{1\nu} + k_{1\mu}k_{3\nu} - k_{3\mu}k_{1\nu} )
\bigg) \partial_\mu^{k_2}\\
\nonumber
&& \hspace{-3cm}
+ \frac 14 \bigg(
k_{1\mu}k_{2\nu} + k_{2\mu}k_{1\nu} - 2 \delta_{\mu\nu} k_{1\rho}k_{2\rho}
\bigg) ( \partial_\mu^{k_3} + \partial_\mu^{k_4} )\Bigg]\\
&& \hspace{-2.5cm}\times \bigg[
\delta^{(n)} ((k_1 \oplus k_2) \ominus (k_3 \oplus k_4))
+\delta^{(n)} ((k_1 \oplus k_2) \ominus (k_4 \oplus k_3))
\bigg]\,.
\label{Feynrulekappa}
\end{eqnarray}
In the above, the sum of $\delta$-functions
represents all mutually different physical situations.

The $\delta$-functions in (\ref{Feynrulekappa}), should in principle  
come from the contraction of fields quite naturally, if the
noncommutative version of the LSZ formalism is applied to our
model. Since such a formalism has not been developed so far, we choose to
follow an approach that kind of combines the standard quantum field theory
consideration (used when treating terms in the Lagrangian linear in $a$ as
small perturbations) with the peculiarities resulting from 
the $\kappa$-deformed statistics of particles. The latter part is realized through embedding a
nonabelian momentum-energy conservation within the 4-point vertex function. 
That approach may seem as a {\it hybrid construction} arising from
trying to move our understanding one step forward towards a
complete quantum theory on noncommutative spaces in general.
In this sense, such an approach can serve as an
intermediate step bridging the gap between the standard quantum field theory and a
complete field theory on $\kappa$-space in a similar way as
for example the semiclassical theory of radiation can be considered as
a cross-over towards the quantum theory of radiation.
In following the described path we have to 
rely in part on intuition, especially  when peculiarities of $\kappa$-space
statistics have to be taken into account. 

The Feynman rule (\ref{Feynrulekappa}) now appears to be consistent
with the energy-momentum conservation that respects 
the $\kappa$-deformed momentum addition rule.
In order to obtain the complete expression for 
the $\delta$-functions appearing in (\ref{Feynrulekappa}),   
we are proceeding in two steps.
First, with the help of (\ref{4+}) and (\ref{7}), up to linear order in $a$,
and with $j,l=3,4;\, j\not= l$, we have:
\begin{eqnarray}
\delta^{(n)} ((k_1 \oplus k_2) \ominus (k_j \oplus k_l))
&=&\frac{\delta^{n}((k_1 \oplus k_2) - (k_j \oplus k_l))}
{{\bigg(1 - a(k_1 \oplus k_2)\bigg)}^{n-1}} 
\label{deltakappa}\\ 
&& \hspace{-1.3cm}
=\frac{\delta^{(n)}((k_1 \oplus k_2)-(k_j \oplus k_l))}
{{\bigg[1-\bigg(a(k_1 + k_2)+(ak_1)(ak_2)-a^2(k_1k_2)
+ {\mathcal O}(a^3)\bigg)\bigg]}^{n-1}}
\nonumber\\
&& \hspace{-1.3cm}
=(1+(n-1)a(k_1+k_2)+ {\mathcal O}(a^2))
\delta^{(n)}((k_1\oplus k_2)-(k_j \oplus k_l)) \,.
\nonumber
\end{eqnarray}

The second step is to compute the $\delta$-functions from (\ref{deltakappa})
in the same way as we computed the one in (\ref{7}):
\begin{equation}
\delta^{(n)}((k_1\oplus k_2)-(k_j \oplus k_l))=\sum_i
\Bigg|\det \Bigg(\frac{\partial {((k_1 \oplus k_2) 
-(k_j \oplus k_l)}_{\mu}}{\partial k_{1\nu}}\Bigg)_{k_1=q_i} 
\Bigg|^{-1}
\delta^{(n)}(k_1-q_i) \,,
\label{6a}
\end{equation}
where we have to sum up over all zeros $q_i$ for 
the expression in the argument of $\delta$-function. 

To proceed, we shall choose specific values for the momenta $k_2=k_3=\ell$, which
we need for the evaluation of the tadpole diagram.
Since there are no zeros for the delta function 
$\delta^{(n)}((k_1 \oplus \ell) - (\ell \oplus k_4))$, the only 
contribution comes from the second combination 
$\delta^{(n)}((k_1\oplus\ell)-(k_4\oplus\ell))$.
In order to perform that computation, 
we start with (\ref{4+}) and  orient the vector $a$
in the direction  of time, $a =(0,...,0,ia_0). $ 
Then due to covariance, the obtained result will also
 be valid for an arbitrary orientation of $a$. Hence
\begin{eqnarray}
\label{Jacobian}
  \det {\left( \frac{\p {((k_{1} \oplus \ell) - (k_{4} \oplus
          \ell))}_{\mu}}{\p k_{1\nu }} \right)}_{k_{1} = k_{4}}
&=&  \left| \begin{array}{ccccc}
  1 & ~ - ia_0 \ell_1 & 
   \cdot \cdot \cdot & ~ - ia_0 \ell_{n-2}
   & ~  - ia_0 \ell_{n-1} \\
 0  & 1+ a\ell &  \cdot \cdot \cdot
 & 0 & 0 \\ 
 \cdot & \cdot &   \cdot \cdot \cdot & \cdot & \cdot \\
\cdot & \cdot &  \cdot \cdot \cdot & \cdot & \cdot \\
 0  &  0 &  \cdot \cdot \cdot 
 & 1 + a\ell & 0 \\
  0 & 0 & \cdot \cdot \cdot
 & 0 &  1 + a\ell \\
\end{array} \right|  
\nonumber\\
&=&  {(1 + a\ell)}^{n-1}.
\end{eqnarray}
Since there is only one zero, at $q_i=k_4$, we find 
\begin{eqnarray}
\hspace{-5mm}
\delta^{(n)}((k_1 \oplus \ell) - (k_4 \oplus \ell))  
=\frac{\delta^{(n)}(k_1 - k_4)}{{(1 + a\ell)}^{n-1}}
= \bigg (1 - (n-1) a\ell  \bigg ) \delta^{(n)}(k_1 - k_4) 
+\mathcal O(a^2) \,,
\label{tadpoleaksi3}
\end{eqnarray}
which gives us the final expression for the $\delta$-function in (\ref{Feynrulekappa})
via (\ref{deltakappa}), up to first order in $a$:
\begin{eqnarray}
\delta^{(n)}((k_1 \oplus \ell) \ominus (k_4 \oplus \ell ))
&=&\bigg(1+(n-1)a(k_1+\ell)\bigg)\frac{\delta^{(n)}(k_1 -k_4)}{{(1+a\ell)}^{n-1}}
+\mathcal O(a^2)
\nonumber\\
&=&(1 + (n-1)ak_1)\delta^{(n)}(k_1  - k_4) + \mathcal O(a^2)\,.
\label{delta-}
\end{eqnarray}
In the above expression the dependences on $\ell$ drop out as we
expected, thus showing the consistency of the {\it hybrid} Feynman rule derivation. 
The remaining factor $(1 + (n-1)ak_1)$ in (\ref{delta-}) 
is due to the $\kappa$-space twisted 
particle statistics of our {\it hybrid approach}.

\subsection{Massive scalar field propagation }
\subsubsection{Tadpole diagram: standard momentum conservation}

\begin{figure}
%\FIGURE{figure}
\begin{center}
\includegraphics[width=70mm]{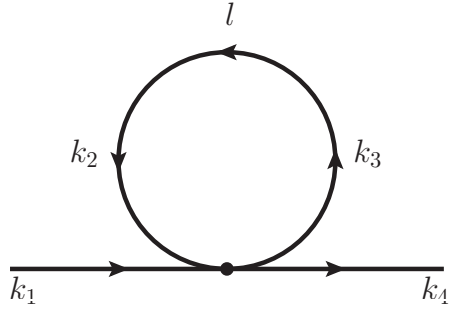}
\end{center}
\caption{Scalar 4-field tadpole}
\label{fig:tadpol}
\end{figure}

In order to compute the tadpole diagram from Fig. \ref{fig:tadpol},
using dimensional regularization technique,
we have to introduce in the action (\ref{actioncomlex}) 
$new$ dimensionfull regularization masses, denoted by $\mu$
for the coupling $\lambda$, and by $\mu'$ for $\xi^2$, respectively.
In accordance with quantum field theory \cite{Casalbuoni}, the
regularization of the $\phi^4$ model requires that they are given in the following form:
\begin{eqnarray}
\lambda_{new}&=&\lambda_{old}(\mu^2)^{\frac{n}{2}-2}
\;\;\; \to \;\;\;(\mu^2)^{2-\frac{n}{2}}\lambda\,,\;\;\;\lambda=\lambda_{new}\,,
\label{lambdamu}\\
\xi_{new}&=&\xi_{old}({\mu'}^2)^{\frac{n}{2}-4}
\;\;\; \to \;\;\;({\mu'}^2)^{4-\frac{n}{2}}\xi\,,\;\;\;\xi=\xi_{new}\,.
\label{ximu}
\end{eqnarray}
Here $\mu'$ is defined in such a way that the new parameters are dimensionless.
%$dim[\xi^2]=dim[mass^4]$, for $n=4$.
%under the loop integral contribution from $\xi$-term in (\ref{actioncomlex}).

In this paper, we shall further restrict our computation only to the
first order contribution of the two-point function ${\Pi}^{a,\xi}_2$ in our 
$\lambda\phi^4$ model, (\ref{actioncomlex}), corresponding to the 
tadpole diagram depicted in Fig. \ref{fig:tadpol}:
\begin{equation}
\Pi^{a,\xi}_2 = 
{\Pi}^{0,0}_2+{\Pi}^{a\ne 0,0}_2+{\Pi}^{0,\xi\ne 0}_2+{\Pi}^{a\ne 0,\xi\ne 0}_2=
\int\frac{d^n k_2}{(2\pi)^n}\frac{d^n k_3}{(2\pi)^n}\,
\tilde\Gamma(k_1,k_2,k_3,k_4;a,\mu) G^{\xi}(k_2,k_3;\mu')\,.
\label{3}
\end{equation} 
In general, the one-loop integral (\ref{3}) produces 
four different contributions, where for $G^{\xi}(k_2,k_3;\mu')$ 
we are using the expanded expression 
(\ref{propagator1}) further on. 
In computing the first two terms from (\ref{3}), we assume momentum conservation, 
$k_1 + k_2 = k_3 + k_4$ and $k_2 = k_3 = \ell$. Thus we have
\begin{eqnarray}
\hspace{-.5cm}
{\Pi}^{a,0}_2=
%{\Pi}^{0,0}_2+{\Pi}^{a\ne 0,0}_2=
%+{\Pi}^{0,\xi\ne 0}_2+{\Pi}^{a\ne 0,\xi\ne 0}_2=  
%T^{0,0}_2+T^{a,0}_2+T^{0,\xi}_2=
\int \frac{d^n\ell}{(2\pi)^n} \;
\tilde\Gamma(k_1,\ell,\ell,k_4;a,\mu)\,\frac{i}{\ell^2+m^2}\,.
\label{tadpole}
\end{eqnarray}
 First, we need
$\tilde\Gamma$ from the Feynman rule (\ref{Feynrule}), 
in accordance with the notations from Fig. \ref{fig:vertex}; that is 
for incoming momenta $k_1$ and outgoing momenta $k_4$ 
we have to replace $k_3 \to -k_3=-\ell$ and $k_4 \to -k_4$. Thus, we have
\begin{eqnarray}
\label{Gamma}
&& \hspace{-1cm}
\tilde\Gamma(k_1,\ell,\ell,k^{out}_4;a,\mu) 
 = i(2\pi)^{n} \mu^{4-n}\;\frac{\lambda}{2}\, 
\Bigg\{1+ a_{\nu}\frac{1}{8}
\bigg[\bigg(k_{4\mu}k_{1\nu} - k_{1\mu}k_{4\nu} \bigg)\partial^{\ell}_{\mu}
\\
\nonumber
\hspace{-1cm}
&-&  
2\bigg({\ell}_{\mu}k_{4\nu} -3\ell_{\nu}k_{4\mu}
+8\delta_{\mu\nu}k_{4\rho}\ell_\rho\bigg)\partial^{k_1}_{\mu}
-2\bigg(\ell_{\mu}k_{1\nu}+k_{1\mu}\ell_{\nu}
-2\delta_{\mu\nu}k_{1\rho}\ell_\rho\bigg)\partial_\mu^{k_4}\\
\nonumber 
\hspace{-1cm}
&+&  
\bigg(\ell_{\mu}(2k_4-k_1)_{\nu}+\ell_{\nu}(2k_4-3k_1)_{\mu} - 
4 \delta_{\mu\nu} (k_4 - k_1)^{\rho}\ell_{\rho} 
 \bigg)\partial^{\ell}_{\mu} \bigg]
\Bigg\} \delta^{(n)}(k_1 - k_4)\,.
\nonumber
\end{eqnarray}
In order to obtain $\tilde\Gamma(k_1,\ell,\ell,k^{in}_4;a,\mu)$ we just have 
to replace $k_4 \to -k_4$ in (\ref{Gamma}).

As a next step, we compute $\Pi^{a,\xi}_2$ straightforwardly.
%with the help of notion for the 
%integral (\ref{tadpole}) as an effective action describing 
%the given one-loop quantum
%process. 
Employing integration 
by parts in (\ref{tadpole}) and
using the dimensional regularization scheme, we obtain
\begin{eqnarray}
\Pi^{0,0}_2&=&-\frac{\lambda}{2}\,I_0 \,,
\label{T0}\\
\Pi^{a\ne 0,0}_2&=&-\frac{\lambda}{2} \Bigg\{ \frac{3}{8} {(aK)}I_0 -
\frac{1}{4}{(aK)}\bigg( {\delta}_{\mu\nu} 
- \frac{a_{\mu}K_{\nu}}{(aK)}\bigg)I_{2,\mu\nu}\Bigg\}\,,
\label{Ta}
\end{eqnarray}
where $K=2k_4-k_1$. The oscillator contribution from the $\xi$-term 
has been obtained by a rude approximation of (\ref{3}). That is, from
\begin{eqnarray}
\hspace{-.5cm}
{\Pi}^{0,\xi\ne 0}_2=
-i\int \frac{d^n\ell}{(2\pi)^n} \;
\tilde\Gamma(k_1,\ell,\ell,k_4;a,\mu)\,
\frac{\xi^2}{(\ell^2+m^2)^2}\partial_{\ell}^2 \delta^{(n)}(\ell)\,,
\label{tadpolexi}
\end{eqnarray}
after integrating by parts, we have found
\begin{eqnarray}
\Pi^{0,\xi\ne 0}_2&=&
-\frac{\lambda}{2}\xi^2\frac{(8n)({\mu'}^2)^{4-\frac{n}{2}}}{m^6}
\label{Txi}\,.
\end{eqnarray}
In the above equations, we have omitted the factor $(2\pi)^n \delta^{(n)}(k_1 - k_4)$ for simplicity.
For $n=4-\epsilon$, we have the well known integrals
\begin{eqnarray}
 I_0&=&(\mu^2)^{2-\frac{n}{2}} 
\int \frac{d^{{n}}\ell}{(2\pi)^n}\frac{1}{\ell^2 + m^2} 
=\frac{m^2}{(4\pi)^2} 
\Bigg[\Bigg(\frac{4{\pi}{\mu}^2}{m^2}\Bigg)^{\frac{\epsilon}{2}}
\Gamma(-1+\frac{\epsilon}{2})\;{{\Bigg]}_{\epsilon \to 0}} 
\nonumber\\ 
&=& \frac{-1}{8\pi^2} m^2   
\Bigg[\frac{1}{\epsilon}+\frac{\psi(2)}{2}+{\rm log}\sqrt{\frac{4\pi\mu^2}{m^2}} +...\Bigg]
\label{I0}\,,\\ 
I_{2,\mu\nu}&=& (\mu^2)^{2-\frac{n}{2}} 
\int \frac{d^n\ell}{(2\pi)^n}\frac{\ell_{\mu}\ell_{\nu}}{(\ell^2+m^2)^2}=
\frac{1}{2} \delta_{\mu\nu} I_0 \,,\;\;\; \delta_{\mu\nu}\delta_{\mu\nu}=n\,,
\label{I2}
\end{eqnarray}
with a simple pole at $\epsilon =0$.
Thus the expression (\ref{I0}) is divergent in the UV cut-off.

The non-vanishing contributions come from the commutative cases, 
that is from (\ref{T0}), and from harmonic oscillator term
(\ref{Txi}) which is finite for finite scalar field mass.
The integrals (\ref{I0}) and (\ref{I2}) for $n=4$ give $\Pi^{a\ne 0,0}_2=0$,
producing in the case $\xi=0$, very well known commutative 
$\lambda\phi^4$ theory result (\ref{T0}). All contributions
proportional to $a$, coming from $\kappa$-Minkowski NC $\phi^4$ theory cancel out,
as one would naively expect by inspecting the vertex weight (\ref{Gamma}). 

Clearly, the one loop computation has to be modified by
anticipating the momentum conservation on $\kappa$-space.
To illustrate that something nonstandard appears in our model (\ref{Feynrule}), 
we start with the general one-loop integral (\ref{3}).
It should be noted that one cannot integrate over $k_3$ using 
the first delta $\delta^{(n)}(k_2 - k_3)$ -- from the propagator -- 
and replace $k_3$ by $k_2$ in the above
expression as it stands, because of the derivative with respect to $k_2$.
So, in a first step of the computation we are using a simple trick 
\begin{eqnarray}
\partial_\mu^{k_1} \delta^{(n)}(k_1 + k_2 - k_3 - k_4) &=& 
\partial_\mu^{k_2} \delta^{(n)}(k_1 + k_2 - k_3 - k_4)\,,
\nonumber\\
\partial_\mu^{k_1} \delta^{(n)}(k_1 + k_2 - k_3 - k_4) &=& 
-\partial_\mu^{k_3} \delta^{(n)}(k_1 + k_2 - k_3 - k_4)\,, \; {\rm etc.,}
\label{del3B}
\end{eqnarray}
and than we rewrite (\ref{Feynrule}) and (\ref{propagator1}) as follows:
\begin{eqnarray}
 &&\tilde\Gamma(k_1,k_2,k_3,k_4;a)G^{\xi}(k_2,k_3) = 
\nonumber\\
&& 
\Bigg\{i(2\pi)^n\frac{\lambda}{4} 
\Bigg[1+a_\nu \bigg(2(-k_{1\mu}k_{2\nu} - k_{2\mu}k_{1\nu} + k_{3\mu}k_{4\nu} +
k_{4\mu} k_{3\nu} + 2 \delta_{\mu\nu} (k_{1\rho}k_{2\rho}-k_{3\rho}k_{4\rho}))
\nonumber\\
&&
+\frac{1}{2}(k_{1\mu}k_{3\nu}-k_{3\mu}k_{1\nu}+k_{2\mu}k_{3\nu}-k_{3\mu}k_{2\nu}
+ k_{1\mu}k_{4\nu} - k_{4\mu}k_{1\nu} + k_{2\mu}k_{4\nu}-k_{4\mu}k_{2\nu})\bigg)
\partial_\mu^{k_1}\Bigg] 
\nonumber\\
&&
\times\delta^{(n)} (k_1 + k_2 - k_3 - k_4)\Bigg\}\bigg[
\frac{i}{k_2^2+m^2} \bigg(1-\frac{\xi^2}{(k_2^2+m^2)}\partial^2_{k_3}\bigg)
\delta^{(n)}(k_2-k_3)\bigg]
\,.
\label{Feynrule2}
\end{eqnarray} 
For the part independent of $\xi$, we obtain after performing 
the integration over $k_3$ in (\ref{Feynrule2}) the following expression:
\begin{eqnarray}
&&\Pi^{a \ne 0, 0}_2= -\frac{\lambda }{8(2\pi)^n} ((ak_1)k_4-(ak_4)k_1)_{\mu}
\bigg[\partial_\mu^{k_1} \delta^{(n)}( k_1 - k_4 )\bigg]  
 \int \frac{d^n k_2}{k_2^2 + m^2}\,,
\label{del6B}
\end{eqnarray}
where $k_1$ and $k_4$ are external momenta. 
This expression is quadratically divergent in the UV cut-off
representing a quantum loop modification of 
the free action (\ref{actioncomlex}), 
which can be nonzero because of the momentum conservation violation at the vertex. 
The next term of (\ref{Feynrule2}), is obtained after partial integration:
\begin{eqnarray}
{\Pi}^{0,\xi\ne 0}_2  =  
- \frac{\lambda\xi^2}{4(2\pi)^n} \bigg[\partial^2_{k_1} \delta^{(n)} (k_1 - k_4)\bigg] 
\int  \; \frac{d^n k_2}{(k_2^2+m^2)^2} \,.
\label{0xi0}
\end{eqnarray}
It is logarithmically divergent in $n=4$ dimensions.
The last term depends on both deformation parameters, $a$ and $\xi$, it reads
\begin{eqnarray}
{\Pi}^{a\ne 0,\xi\ne 0}_2 & = & 
- \frac{\lambda \xi^2}{8(2\pi)^n} a_\nu \bigg[
\bigg( \partial^2_{k_1} \partial_\mu^{k_1} \delta^{(n)} (k_1 - k_4)\bigg) 
(k_{1\mu}k_{4\nu}-k_{4\nu}k_{1\mu})  
\nonumber\\
&& -4 %\frac{\lambda \xi^2}{2(2\pi)^n} a_\nu 
\bigg( \partial^{k_1}_\sigma \partial_\mu^{k_1} \delta^{(n)} (k_1 - k_4)\bigg)
\bigg(2(\delta_{\mu\sigma}k_{4\nu} 
+ \delta_{\nu\sigma}k_{4\mu} - 2 \delta_{\mu\nu} k_{4\sigma})
\nonumber\\
&& \hspace{1cm}
+ \frac12 (\delta_{\nu\sigma}k_{1\mu}-\delta_{\mu\sigma}k_{1\nu})\bigg) \bigg]
\int \frac{d^n k_2}{(k_2^2+m^2)^2}\,, 
\label{a0xi0}
\end{eqnarray}
which is also logarithmically divergent in $n=4$ dimensions. 
The one-loop corrections \eqref{0xi0}-\eqref{a0xi0} 
have to be included in the action as counterterms. 
Their structure is different from the tree-level terms, 
and therefore the tree-level action is not stable under 1-loop quantum corrections. 
For this reason, we have to question the approximations we have employed: 
namely the expansion of the action up to first order 
in the deformation parameter $a$ and the expansion of 
the propagator up to first order in $\xi$. 
The results \eqref{T0} and \eqref{Ta} (as well as the result (\ref{0a0}) 
that will follow shortly) are obtained
directly from \eqref{tadpole}. 
We assume that \eqref{tadpole} replaces the expression \eqref{3} for $\xi=0$.
As for \eqref{Txi} (and \eqref{a,x-i}), 
it is a rude estimate of what the $\xi$ term would produce.
 It is obtained by making  approximations in \eqref{3}, by approximating/adjusting
 the form of the propagator \eqref{propagator1}
 and by using the $\kappa$-deformed addition/subtraction
  of momenta instead of the commutative one. The difference between \eqref{Txi} 
  and \eqref{a,x-i} is
  that unlike the former expression, the latter one is obtained by making use of the 
$\kappa$-deformed addition/subtraction rule for momenta. 
Due to the results \eqref{del6B}-\eqref{a0xi0} 
we obviously stumbled across the momentum nonconservation. 
Such results seem to favor our {\it hybrid} approach.

\subsubsection{Tadpole diagram: $\kappa$-deformed momentum conservation}

In the following computation of the 
tadpole diagram from Fig. \ref{fig:tadpol}, 
we fully implement the {\it hybrid approach}.
That is the notion that standard momentum conservation is not satisfied,
i.e. we use induced momentum conservation on $\kappa$-space represented 
within the $\delta$-functions in (\ref{delta}). 
However, in accordance with our {\it hybrid approach},
at the end undeformed momentum conservation has to be applied. 
The general one-loop integral (\ref{3}) 
can be roughly reduced to two terms, namely (\ref{tadpole}) and (\ref{tadpolexi}).  

In the next step, we are applying integration by parts, which is possible due to 
the notion that the remaining integral in (\ref{3}) is an effective action
of the given one-loop quantum process. 
This of course plays an essential role in our {\it hybrid approach}.
Performing the computation
of all terms in the tadpole one-loop integral (\ref{tadpole})
with Feynman rule (\ref{Feynrulekappa}) and $\delta$-function (\ref{delta-}) and
for an arbitrary number of dimensions $n$, 
we obtain the following first order result:
\begin{eqnarray}
\Pi^{0,0}_2 + \Pi^{a\ne 0, 0}_2 &=& -\frac{\lambda}{2} 
\bigg[(1+(n-1)ak_1)+\frac{1}{2}(1-\frac{n}{4})(ak_1-2ak_4)\bigg]\,I_0 
\label{0a0}\\
\Pi^{0,\xi\ne 0}_2 + \Pi^{a\ne 0,\xi\ne 0}_2
&=&-\frac{\lambda}{2}\xi^2\frac{8n({\mu'}^2)^{4-\frac{n}{2}}}{m^6}
(1+(n-1)ak_1)\,.
\label{a,x-i}
\end{eqnarray}
The first terms in (\ref{0a0}) and (\ref{a,x-i}) for $n =4$ correspond to the results 
(\ref{T0}) and (\ref{Txi}) from the previous subsection. 
From the above formulas it is clear that there exist 
non-vanishing contributions even for $n=4$ dimensions.
They are arising 
via the $\kappa$-deformed momentum conservation rule, entering through
the deformed $\delta$-function (\ref{delta-}) in the {\it hybrid} Feynman rule 
(\ref{Feynrulekappa}), and from the harmonic oscillator 
term in the action (\ref{actioncomlex})
via the modified propagator (\ref{propagator1}).

For $n=4-\epsilon$, we obtain a modified expression 
for the tadpole in Fig. \ref{fig:tadpol}
in the limit $\epsilon \to 0$, where the $1/\epsilon$-divergence is explicitly isolated. 
For conserved external momentum in accordance with (\ref{delta-}), 
i.e. for $k_1=k_4\equiv k$, 
from (\ref{I0}), (\ref{0a0}) and (\ref{a,x-i}) we finally have, 
\begin{equation}
{\Pi}^{a,\xi}_2=\frac{\lambda m^2}{32\pi^2} (1+3ak)   
\Bigg[\frac{2}{\epsilon}+\psi(2)+{\rm log}\frac{4\pi\mu^2}{m^2}
-\frac{9}{4}\frac{ak}{1+3ak}
 -(4-\epsilon)128\pi^2\frac{\xi^2 {\mu}^{4-\epsilon}}{m^8}\Bigg]\,,
\label{n=4}
\end{equation}
where for simplicity we have used $\mu'=\mu$,
and in the $\xi$-term we retain the explicit $\epsilon$-dependence
in order to keep dimensional and/or limiting procedure under control. 
The parts above represent modifications of the scalar field self-energy
and explicitly depend on the regularization parameter, 
the mass of the scalar field, 
the magnitude of the translation invariance breaking, and it
contains a correction $ak$ due to
the dependence on the energy $|k|$,
where actual scalar field self-energy modification occurs.  
The dependence of (\ref{n=4}) on the $\kappa$-deformation parameter $a$
enters explicitly, as we expected. Note that there
is no need to do renormalization at the point $(1+3ak) \to 0$.

The result (\ref{n=4}) is next discussed in the framework of Green's functions.
Generally we know that by summing all the 1PI contributions,
for the modifications of the full free propagator (\ref{propagator}),
we get the following expression for 
the {\it two-point connected} Green's function \cite{Casalbuoni}
\begin{eqnarray}
G_{(c,2)}^{a,\xi}(k_1,k_4)&=&%(2\pi)^n\delta^{(n)}(k_1-k_4)
\bigg[\frac{i}{k_1^2+m_1^2}+
\frac{i}{k_1^2+m_1^2}{\Pi}^{a,\xi}_2\frac{i}{k_1^2+m_1^2} +...\bigg]\,,
\label{Gc,2}
\end{eqnarray}
where, symbolically, $m_1^2=m^2+\xi^2\partial^2_{k_1}\delta^{(n)}(k_1-k_4)$ 
represents the redefined mass.  
As an illustration we resume the above series in the limit $\xi\to 0$:
\begin{eqnarray}
G_{(c,2)}^{a,\xi}(k_1,k_4)
 &{}_{\longrightarrow}^{\xi\to 0}&\, 
 (2\pi)^n\delta^{(n)}(k_1 - k_4)\bigg[\frac{i}{k_1^2+m^2-{\Pi}^{a,0}_2}\bigg]\,.
\label{T2}
\end{eqnarray}

The genuine $1/\epsilon$ divergence in ${\Pi}^{a,\xi}_2$,
can only be removed by introducing the following counter term $\delta m^2$:
\begin{equation}
\delta m^2 \tilde \phi^*(k) \tilde \phi(k) = \frac{\lambda m^2}{32\pi^2}    
\Bigg[2(1+3ak)\frac{1}{\epsilon} 
+f\bigg(\frac{4-\epsilon}{2},\frac{\mu^2}{m^2},ak,\frac{\xi^2}{m^4}\bigg)
\Bigg]\tilde \phi^*(k) \tilde \phi(k)\,,
\label{Rn=4}
\end{equation}
where $f$ denotes an arbitrary dimensionless function, 
which is fixed by the renormalization conditions.
Adding the contribution of the above counterterm to the previous expression (\ref{Gc,2})
results during the renormalization procedure  in 
the shift $m^2 \to m^2 + \delta m^2$  in (\ref{Gc,2})/(\ref{T2}), 
thus leading to
\begin{eqnarray}
{\tilde G}_{(c,2)}^{a,\xi}(k_1,k_4)&=&
\Bigg[G_{(c,2)}^{a,\xi}(k_1,k_4)+
G_{(c,2)}^{a,\xi}(k_1,k_4)(-\delta m^2)G_{(c,2)}^{a,\xi}(k_1,k_4) +...\Bigg]
%_{\xi\to 0}
\nonumber\\
 &{}_{\longrightarrow}^{\xi\to 0}& (2\pi)^n\delta^{(n)}(k_1-k_4)
\bigg[\frac{i}{k_1^2+m^2+\delta m^2  
 -{\Pi}^{a,0}_2}\bigg]\,,
\label{T2m}
\end{eqnarray} 
where ${\tilde G}_{(c,2)}^{a,\xi}$ denotes the Green's function 
with the contribution from the counterterm incorporated.

However, since ${\Pi}^{a,\xi}_2$ was computed for the 
expanded free propagator (\ref{propagator1}), it is consistent to
compute the {\it two-point connected} Green's function
under the same approximation.
After the resummation of (\ref{Gc,2}) and using (\ref{propagator1}), we obtain
\begin{eqnarray}
G_{(c,2)}^{a,\xi}(k_1,k_4)
=%(2\pi)^n\delta^{(n)}(k_1 - k_4)
\frac{G^{\xi}}{1-G^{\xi}\;{\Pi}^{a,\xi}_2} \,.
\label{Gexp}
\end{eqnarray}
In order to identify the proper counterterm for the above expression, we resume the series in (\ref{T2m}),
but with replacement (\ref{Gc,2}) $\to$ (\ref{Gexp}),
and the full free propagator $i/(k_1^2+m_1^2)$ replaced 
with the expanded one (\ref{propagator1}), giving:
\begin{eqnarray}
{\tilde G}_{(c,2)}^{a,\xi}(k_1,k_4)
=%(2\pi)^n\delta^{(n)}(k_1 - k_4)
\frac{G^{\xi}}{1+G^{\xi}(\delta m^2-{\Pi}^{a,\xi}_2)} \,.
\label{tilGexp}
\end{eqnarray}
In (\ref{tilGexp}) $\delta m^2$ is a generic quantity. 
This is due to the fact that expression (\ref{n=4}) also contains finite parts.
The requirement $\delta m^2={\Pi}^{a,\xi}_2$ removes the infinity.
Thus we have
\begin{eqnarray}
{\tilde G}_{(c,2)}^{a,\xi}(k_1,k_4)
=\frac{G^{\xi}}{1+G^{\xi}\frac{\lambda m^2}{32\pi^2}
\bigg[f-(1+3ak)\bigg(\psi(2)+{\rm log}\frac{4\pi\mu^2}{m^2}
-\frac{9}{4}\frac{ak}{1+3ak} -512\pi^2\frac{\xi^2 {\mu}^{4}}{m^8}\bigg)\bigg]} \,.
\nonumber\\
\label{tilGexpf}
\end{eqnarray}

Precise extraction and removal of the genuine UV divergence is performed next via 
(\ref{Rn=4})
in the context of the analysis of 
${\tilde G}_{(c,2)}^{a,\xi}(k_1,k_4)$ for different energy regimes; 
that is from low energy to extremely high -Planck scale- energy propagation.

There is a very interesting property of expression (\ref{n=4}) at extreme energies. 
Namely, there exists a term $(1+3ak)$ which for $(1+3ak \to 0)$ tends to zero linearly.
For low energies and/or small $\kappa$-deformation $a$, i.e. 
for $ak = 0$, (equivalent to $a=0$), which is far away from the point $(1+3ak=0)$, 
this is not the case. \\

\noindent
{\it Low energy limit}\\
Using the finite combination $(\delta m^2 - \P^{0,0}_2)$ 
for low energy ($ak = 0$), and at order $\lambda$ 
\begin{equation}
\delta m^2-{\Pi}^{0,0}_2=\frac{\lambda m^2}{32\pi^2}    
\Bigg[f-\psi(2)-{\rm log}\frac{4\pi\mu^2}{m^2}\Bigg]\,,
\label{Rn4}
\end{equation}
from (\ref{n=4}) and (\ref{tilGexpf}), 
%in addition to removing spurious term
%$\xi^2\partial^2_{k}\delta^{(4)}(k)$, 
we get
\begin{eqnarray}
{\tilde G}_{(c,2)}^{0,\xi}(k_1,k_4)
=\frac{G^{\xi}}{1+G^{\xi}\frac{\lambda m^2}{32\pi^2}
\bigg(f-\psi(2)-{\rm log}\frac{4\pi\mu^2}{m^2}
 +512\pi^2\frac{\xi^2 {\mu}^{4}}{m^8}\bigg)} \,,
\nonumber\\
\label{tildeGc2}
\end{eqnarray}
while in the case $\xi=0$
\begin{eqnarray}
{\tilde G}_{(c,2)}^{0,0}(k_1,k_4)
=\frac{(2\pi)^4\delta^{(4)}(k_1-k_4)}{k_1^2+m^2\bigg(1+\frac{\lambda}{32\pi^2}
\bigg[f-\psi(2)-{\rm log}\frac{4\pi\mu^2}{m^2}\bigg]\bigg)} \,.
\nonumber\\
\label{tildeGc200}
\end{eqnarray}
This expression has a pole in Minkowski space, and we can define 
the renormalization condition by requiring
that the inverse propagator at the physical mass is $k_1^2 + m^2_{phys/low}$.
This choice, in the case $\xi^2=0$, 
determines uniquely the sum of the residual terms in (\ref{Rn4}),
which is in accordance with the result from commutative $\phi^4$ theory \cite{Casalbuoni}.\\

\noindent
{\it Planckian energy limit}\\
At the limiting point $(1+3ak\to 0)$, which corresponds 
to extreme energy propagation $|k|$,
where the components of the $\kappa$-deformation parameter $a_\mu$ are extremely small,
of Planck length order, the divergence in (\ref{n=4}) gets removed
{\it under the reasonable assumption that $(1+3ak)$ tends to zero with 
the same speed as $\epsilon$ does}. 
That is, in the Planckian propagation energy limit
\begin{equation}
\frac{(1+3ak)\to 0}{\epsilon \to 0} \longrightarrow {\cal O}(1)\,,
\label{aklimit}
\end{equation}
the $1/\epsilon$ and $ak$ terms from (\ref{n=4}) do contribute.
  
Assuming that our $\kappa$-noncommutativity is spatial,
$a_{\mu}=(\vec a,0)$, and using the momentum
along the third axis $k_{\mu}=(0,0,E,iE)$,
i.e. for $ak=Ea_3$, Eq. (\ref{n=4}) in the Planckian propagation energy 
limit (\ref{aklimit}) gives 
\begin{equation}
\Pi^{a,0}_2\,\bigg|_{(3E a_3 +1\to 0)} 
\longrightarrow\,\frac{\lambda m^2}{32\pi^2}\;
\bigg[2-\frac{9}{4}ak\bigg]_{(3E a_3 +1\to 0)}\,,
\label{ak=1/3}
\end{equation}
producing the following modified Green's function (\ref{T2}):
\begin{equation}
\Pi^{a,0}_2\,\bigg|_{(3E a_3 +1\to 0)}\to 
\frac{\lambda m^2}{32\pi^2}\,\frac{11}{4}\,
\Rightarrow\;
{\tilde G}_{(c,2)}^{a,0}(k_1,k_4)\bigg|_{(3E a_3 +1\to 0)}\simeq 
\frac{i(2\pi)^4\delta^{(4)}(k_1-k_4)}
{k_1^2+m^2\bigg(1-\frac{\lambda}{32\pi^2}\frac{11}{4}\bigg)}\,.
\label{11/4}
\end{equation}
At exact zero-point $3E a_3+1=0$ however, from (\ref{n=4}) 
we obtain a different result 
\begin{equation}
\Pi^{a,0}_2\,\bigg|_{(3E a_3 +1= 0)}= 
\frac{\lambda m^2}{32\pi^2}\,\frac{3}{4}\,\;\;
\Rightarrow\;\;
{\tilde G}_{(c,2)}^{a,0}(k_1,k_4)\bigg|_{(3E a_3 +1= 0)}\simeq 
\frac{i(2\pi)^4\delta^{(4)}(k_1-k_4)}
{k_1^2+m^2\bigg(1-\frac{\lambda}{32\pi^2}\frac{3}{4}\bigg)}\,.
\label{3/4}
\end{equation}

For time-space noncommutativity with the time component of 
$\kappa$-deformation parameter $ia_0$ being also
of order Planck length and vanishing space component, 
i.e. for $ak=-Ea_0$, Eq. (\ref{n=4}) in the Planckian energy 
limit (\ref{aklimit}) yields the result:
\begin{equation}
\Pi^{a,0}_2\,|_{(3E a_0 -1\to 0)} 
\longrightarrow\,\frac{\lambda m^2}{32\pi^2}\;
\bigg[2-\frac{9}{4}ak\bigg]_{(3E a_0 -1\to 0)}\,,
\label{3/4/}
\end{equation} 
equivalent to (\ref{ak=1/3}), 
leading to final results which are identical to (\ref{11/4}) and (\ref{3/4}).

The existence of a linear type of limit $(1+3ak \to 0)$ which removes the genuine UV
divergence $1/\epsilon$ is a new, previously unknown feature of
NC $\kappa$-Minkowski $\phi^4$ theory at linear order in $a$.
The mass shift is of a fixed value (\ref{ak=1/3}),  
independent of the function $f$, but dependent on the energy $|k|$
and the $\kappa$-deformation parameter $a$. 
Above expressions give $\kappa$-deformed dispersion relations, 
thus, (\ref{ak=1/3})/(\ref{3/4/}) and (\ref{11/4})/(\ref{3/4}) represent 
birefringence \cite{Abel:2006wj,Buric:2010wd} of the massive scalar field mode. 
Namely, neither of the finite terms in (\ref{tilGexpf})-(\ref{tildeGc200})
do not contribute in the limit $(1+3ak) \to 0$, and the possibility of
their internal cancellation is diminished. 
In this way, the birefringence of the massive scalar field mode
arises as a genuine effect. 
The above inverse propagator determines the physical mass $m^2_{phys/Planck}$
at respected Planck scale energies.

\section{Discussion and conclusion}

Let us discuss the final outcome of 
the above constructed $\phi^4$ scalar field theory
on the NC $\kappa$-deformed spacetime
and describe the main results of this work.

First, we point out  the most important new results:\\
(i) The integral measure problems on $\kappa$-Minkowski spacetime 
are avoided by the introduction of  
the new ${\star}_h$-product (\ref{newstarproduct}),
which naturally absorbs the measure function due 
to the hermitian realization (\ref{2.10H}).  \\
\noindent
(ii) The trace-like property (the integral identity 
(\ref{property})) is valid for $\kappa$-deformed spaces, but only if one
is dealing with the hermitian realization (\ref{2.10H}), as one should
because only hermitian realizations have a physical meaning.
Due to the integral identity (\ref{property}), the only deformation with respect to the
standard scalar field action comes from the interaction term in (\ref{actionnew}).\\
\noindent
(iii) The action (\ref{actioncomlex}) includes a 
harmonic type of interaction term and is expanded up to first order in the 
deformation parameter $a$, 
producing an effective theory on commutative spacetime.
Despite $\kappa$-deformation a mixture with the $\xi^2$ term in (\ref{actionnew})
via the ${\star}_h$ product, at first order in the $\kappa$-deformation parameter these two
features of our model decouple completely in (\ref{actioncomlex}).
The action (\ref{actioncomlex}) further produces modified 
equations of motion (\ref{eom2}) 
and conserved deformed currents (\ref{current})
due to the invariance under an internal symmetry.
The above properties are very welcome.\\
\noindent
(iv) The truncated $\kappa$-deformed action (\ref{actioncomlex})
does not possess the celebrated quantum effect of UV/IR mixing \cite{Grosse:2005iz}.
The lack of UV/IR mixing is a general feature of most of 
the NC gauge field theories expanded in terms of the deformation parameter.
However, resummation of the expanded action could in principle restore
the nonperturbative character of the model 
(see for example \cite{Horvat:2011iv,arXiv:1109.2485,arXiv:1111.4951}), thus restoring
the presence of UV/IR mixing in quantum loop computations.
Under such circumstances UV/IR mixing would help
in determining what the UV theory might be, that is, it would help to
determine the UV completeness of the theory.
It is important to note that UV/IR mixing connects 
NC gauge theories with Holography via UV and IR cutoffs, 
in a model independent way \cite{Horvat:2010km}. Both, 
UV/IR mixing and Holography, represent important windows to quantum
gravity phenomena \cite{Horvat:2010km,Cohen:1998zx,Szabo:2009tn}.\\
\noindent
(v) Next, we discuss the result of the contribution of the tadpole diagram  
to the propagation
and/or self-energy of our scalar field $\phi$ 
for arbitrary number of dimensions $n$, depicted in Fig. \ref{fig:tadpol}, 
as a function of $\kappa$-deformed momentum conservation law. This deformed conservation law originates from
the deformed statistics on $\kappa$-Minkowski spacetime.
Thus our approach represents kind of a {\it hybrid approach} of standard QFT and NCQFT, involving $\kappa$-deformed $\delta$-functions
in the Feynman rules.\\
\noindent
(vi) When we worked with the standard conservation of momenta
(standard addition of momenta) and undeformed $\delta$-function, the contributions to 
the tadpole diagram to the first order in $a$ are zero. 
The deformed $\delta$-function can be written in terms 
of a leading term plus corrections in $a$. 
Since the Feynman rule (\ref{Feynrule}) has already terms linear in $a$, 
we have to retain only the zeroth order term in
the modified $\delta$-function, because otherwise we get 
terms of quadratic and higher orders in $a$ 
(and this is not what we are interested in). 
The only term where we need to take into account corrections linear in $a$
for the $\delta$-function is the leading order term 
in the Feynman rule (\ref{Feynrule}). However, this term 
vanishes due to the integration over the loop momentum.
Analyzing (\ref{del6B})-(\ref{a0xi0}), we recognize that 
something nonstandard appears in the model 
for the tadpole integral at first order in $\lambda$ and $\kappa$-deformation $a$.\\
\noindent
(vii) It appears that in the computation of the tadpole diagram integrals 
for $n=4$, all contributions linear in $a$ cancel each other automatically. 
For $n\not= 4$ dimensions, the same contributions linear in $a$ become nonzero,
regardless which momentum conservation rule is applied. 
However, the harmonic oscillator term from the action (\ref{actioncomlex})
modifies the mass term in the free propagator (\ref{propagator}), thus 
producing an additional contribution (\ref{Txi}) from the tadpole in Fig. \ref{fig:tadpol}.
The propagation of the scalar field $\phi$ in $n=4$ dimensions also receives a modification from
$\kappa$-deformation at linear order in the deformation parameter $a$, and it also receives a 
contribution due to the oscillator term.\\
\noindent
(viii) In the final computation of the tadpole diagram depicted in Fig. \ref{fig:tadpol}, we fully implement the notion
of our {\it hybrid approach}, that means that standard momentum conservation is not explicitly satisfied,
and we have to use the momentum conservation on $\kappa$-spaces given by (\ref{ominus}), 
while at the end of the computation the undeformed momentum conservation has to be applied. 
We have found non-vanishing contributions even for $n=4$ dimensions.
They are arising from the harmonic oscillator term in the action (\ref{actioncomlex})
via the modified propagator (\ref{propagator1}), 
and the $\kappa$-deformed momentum conservation rule which enters through
the deformed $\delta$-function in the {\it hybrid} Feynman rule (\ref{Feynrulekappa}).
We have found the fully modified expression for the tadpole in Fig. \ref{fig:tadpol}
in the limit $\epsilon \to 0$, where the genuine $1/\epsilon$ (UV) 
divergence is explicitly isolated. 
For conserved external momenta, i.e. for $k_1=k_4\equiv k$, we obtain 
the two-point function (\ref{n=4}),
where the finite parts represent the modification 
of the scalar field self-energy $\Pi^{a,\xi}_2$
and depend explicitly on the regularization parameter $\mu^2$, 
the mass of the scalar field $m^2$, and
the magnitude of the breaking of translational invariance $\xi^2$. The most important is that 
(\ref{n=4}) contains the finite correction $ak$
due to the deformed statistics on $\kappa$-Minkowski spacetime,
thus, via $ak$ term we obtain an explicit dependence on 
the scale of propagating energy involved $|k|=E$, 
and the $\kappa$-deformation parameter $a$, as we expected.\\
\noindent
(ix) The two-point function (\ref{n=4}) is next applied in the framework 
of {\it two-point connected} Green's functions
for three energy regimes, that is for low energies, for Planck scale energies,
and for intermediate energies, respectively.
For low energy scale and/or small $\kappa$-deformation $a$, i.e. 
for $ak\simeq 0$, which is far away from the point $(1+3ak=0)$, 
the dependence of the two-point Green's function on the deformation parameter completely drops out (\ref{tildeGc2}). 
The genuine UV divergence in (\ref{n=4}) 
has been removed by subtracting the counterterm $\delta m^2$ (\ref{Rn=4}),
from the previous contribution (\ref{Gc,2}),
or through shifting $m^2$ into $(m^2 + \delta m^2)$ in (\ref{T2}).
In this case, the mass shift (\ref{tildeGc2}) could increase or decrease $m^2$
depending on the function $f$ and/or the parameter $\xi^2$. \\ 
\noindent
(x) For energies within the limits $\frac{-1}{3a_3} \ll E \ll 0$ (or equivalently
$\frac{1}{3a_0} \gg E \gg 0$), the full expression (\ref{n=4}), 
with mass counterterm (\ref{Rn=4}) has to be used
in determining the Green's function (\ref{T2m}). In that particular case,
the harmonic oscillator term in (\ref{n=4}) will also 
give a non-negligible contribution. It's coupling $\xi^2$
could be in principle determined by higher order contributions 
to the Green's function. This is certainly an issue to be addressed
in future work.  \\
\noindent
(xi) At Planckian energy scale, due to the existence of a
linear type of limits $(1+3ak) \to 0$, we have new situation and,
distinguish two cases. They are both new, previously unknown features of
$\kappa$-deformed $\phi^4$ theory expanded up to linear order in the deformation parameter.
In the first case we have the limit (\ref{aklimit}) which produces
the self-energy and/or modified Green's function (\ref{11/4}).\\
\noindent
(xii) Second case, where $1+3E a_3=0$, 
represents in fact a genuine type of zero-point which  
exactly removes the UV divergence,
yielding the  self-energy and/or modified Green's function (\ref{3/4}).
In both cases the mass term is shifted in the same direction, 
({\it the same sign!}), but for a different amount, $+11/4$ versus $+3/4$, respectively.
Or more precisely we can say that the mass shift during the limiting process  
$(3E a_3 +1\to 0)$ drops from the value proportional to 
$+11/4$ 
%the exact value proportional 
to $+3/4$.\\ 
\noindent
(xiii) The results (\ref{11/4}) and (\ref{3/4}) are the same for two different
choices of $\kappa$-noncommutativity, (with appropriate choice of referent
system for momentum $k_{\mu}$), i.e. for $a_{\mu}=(0,0,a_3,0)$ and 
$a_{\mu}=(0,0,0,ia_0)$, respectively, since (\ref{ak=1/3}) and (\ref{3/4/}) are equivalent.\\
\noindent
(xiv) At Planckian propagation energy scale $E \simeq \frac{-1}{3a_3}$,  
the contribution of the tadpole in Fig. \ref{fig:tadpol} tends to a
finite fixed value, between (\ref{11/4}) and (\ref{3/4}). 
Due to the effects of $\kappa$-Minkowski statistics, this depends on the direction of the 
propagation, the Planckian propagation energy and the deformation parameter $a$.
In this way (\ref{ak=1/3})/(\ref{3/4/}) and (\ref{11/4})/(\ref{3/4})   
represent $\kappa$-deformed dispersion relations and 
produce a birefringence effect \cite{Abel:2006wj,Buric:2010wd} of 
the massive scalar field mode, 
which arises as a genuine effect at first order in $a$. It is similar to the birefringence effect of the 
chiral fermion field in truncated Moyal $\star$-product theories \cite{Buric:2010wd}.\\  
\noindent
(xv) Considering full renormalization, besides the $\delta m^2$ counterterm, 
the other divergent parts have to be added as
counterterms to the action (\ref{actioncomlex}) as well: 
\begin{equation} 
\int d^4 x\,({\cal L} + {\cal L}_{ct}) = S[\phi_B, m_B, \lambda_B, \xi_B, a_B]\,,
\nonumber
\end{equation}
where index $B$ denotes bare quantities and it is assumed that the counterterms can be absorbed by redefinition of the already existing parameters.
That would include analysis of $4-point$ one-loop contributions, 
counterterms $(\mu^2)^{2-\frac{n}{2}}\delta\lambda$ and 
$({\mu}^2)^{4-\frac{n}{2}}\delta\xi$,
as well as 2-loop expansion for the 2-point Green's function with
insertion of counterterms in multi-loop diagrams.
%Certainly, the full analysis of the renormalization group equations 
%is also under the same schedule.
However, the full renormalization of our action (\ref{actioncomlex}) 
is anyhow beyond the scope of this paper and it is planned for our next project. 

%\noindent
Regarding the effects of statistics according to the described arguments, 
it is important to repeat that within first order 
in the deformation parameter $a$, {\it effects of the deformed statistics
on $\kappa$-Minkowski do appear in our hybrid model as
semiclassical/hybrid behavior of the first order quantum effects,
thus showing birefringence of the massive scalar field mode.}
We believe that this property of such a constructed model, is of 
importance for further possible research towards quantum gravity. 
At higher orders in $a$ the
matter would become growingly interesting and complicated.

\noindent
%{\bf Acknowledgment}\\
\acknowledgments
We would like to thank A. Andra\v si, A. Borowiec, H. Grosse, J. Lukierski, 
V. Radovanovi\' c and J. You for fruitful discussions.
We would like to express our especial gratitude to J. Lukierski and A. Borowiec
for careful reading of the manuscript and a number of valuable remarks which we 
incorporated into the final version of this manuscript.
J.T. acknowledges the support from ESI during his stay in Vienna, and W. Hollik
during his stay at the MPI Munich. We thank G. Duplan\v ci\' c for drawing graphs.
This work was supported by the Ministry of Science and Technology of
the Republic of Croatia under contract No. 098-0000000-2865 and 098-0982930-2900.
% The work of J.T. is supported in part by EU (HEPTOOLS) 
% project under contract No. MRTN-CT-2006-035505. 

%%%%%%%%%%%%%%%%%%%%%%%%%%%%%%% bibliography%%%%%%%%%%%%%%%%%%%%%%%%%%%%%%%
 %%%%%%%%%%%%%%%%%%%%%%%%%%%%%%%%%%%%%%%%%%%%%%%%%%%%%%%%%%%%%%%%%%%%%%%%%


\begin{thebibliography}{999}

%\cite{AmelinoCamelia:2003xp}
\bibitem{AmelinoCamelia:2003xp}
  G.~Amelino-Camelia, L.~Smolin and A.~Starodubtsev,
  {\it Quantum symmetry, the cosmological constant and Planck scale
  phenomenology,}
  Class.\ Quant.\ Grav.\  {\bf 21} (2004) 3095
  [arXiv:hep-th/0306134].

%\cite{Freidel:2003sp}
\bibitem{Freidel:2003sp}
  L.~Freidel, J.~Kowalski-Glikman and L.~Smolin,
  {\it 2+1 gravity and doubly special relativity,}
  Phys.\ Rev.\  D {\bf 69} (2004) 044001
  [arXiv:hep-th/0307085.

%\cite{Freidel:2005bb}
\bibitem{Freidel:2005bb}
  L.~Freidel and E.~R.~Livine,
  {\it Ponzano-Regge model revisited. III: Feynman diagrams and effective  field
  theory,}
  Class.\ Quant.\ Grav.\  {\bf 23} (2006) 2021
  [arXiv:hep-th/0502106].

%\cite{Freidel:2005me}
\bibitem{Freidel:2005me}
  L.~Freidel and E.~R.~Livine,
  {\it Effective 3d quantum gravity and non-commutative quantum field theory,}
  Phys.\ Rev.\ Lett.\  {\bf 96} (2006) 221301
   [arXiv:hep-th/0512113].

%\cite{Freidel:2005ec}
\bibitem{Freidel:2005ec}
  L.~Freidel and S.~Majid,
  {\it Noncommutative Harmonic Analysis, Sampling Theory and the Duflo Map in 2+1
  Quantum Gravity,}
  Class.\ Quant.\ Grav.\  {\bf 25} (2008) 045006
  [arXiv:hep-th/0601004].

%\cite{Lukierski:1991pn}
\bibitem{Lukierski:1991pn}
  J.~Lukierski, H.~Ruegg, A.~Nowicki and V.~N.~Tolstoi,
  {\it Q deformation of Poincare algebra,}
  Phys.\ Lett.\  B {\bf 264} (1991) 331.
  
%\cite{Lukierski:1992dt}
\bibitem{Lukierski:1992dt}
  J.~Lukierski, A.~Nowicki and H.~Ruegg,
  {\it New quantum Poincare algebra and k deformed field theory,}
  Phys.\ Lett.\  B {\bf 293} (1992) 344.
  %%CITATION = PHLTA,B293,344;%%
  
%\cite{Majid:1994cy}
\bibitem{Majid:1994cy}
  S.~Majid and H.~Ruegg,
 {\it Bicrossproduct structure of $\kappa$ Poincar\'e group and noncommutative
  geometry,}
  Phys.\ Lett.\  B {\bf 334} (1994) 348
  [arXiv:hep-th/9405107].
  %%CITATION = PHLTA,B334,348;%  
  
\bibitem{Zakrzewski:1994}
S. Zakrzewski, {\it Quantum Poincar\'{e} group related to 
the $\kappa$-Poincar\'{e} algebra},
J. Phys. A: Math. Gen {\bf 27} (1994) 2075.
  
%\cite{Lukierski:1993wx}
\bibitem{Lukierski:1993wx}
  J.~Lukierski, H.~Ruegg and W.~J.~Zakrzewski,
  {\it Classical Quantum Mechanics Of Free $\kappa$ Relativistic Systems,}
  Annals Phys.\  {\bf 243} (1995) 90
  [arXiv:hep-th/9312153].

%\cite{AmelinoCamelia:2000ge}
\bibitem{AmelinoCamelia:2000ge}
  G.~Amelino-Camelia,
  {\it Testable scenario for relativity with minimum-length,}
  Phys.\ Lett.\  B {\bf 510} (2001) 255
  [arXiv:hep-th/0012238].

%\cite{AmelinoCamelia:2000mn}
\bibitem{AmelinoCamelia:2000mn}
  G.~Amelino-Camelia,
  {\it Relativity in space-times with short-distance structure governed by an
  observer-independent (Planckian) length scale},
  Int.\ J.\ Mod.\ Phys.\  D {\bf 11} (2002) 35
  [arXiv:gr-qc/0012051].

%\cite{Magueijo:2001cr}
\bibitem{Magueijo:2001cr}
  J.~Magueijo and L.~Smolin,
  {\it Lorentz invariance with an invariant energy scale,}
  Phys.\ Rev.\ Lett.\  {\bf 88} (2002) 190403
  [arXiv:hep-th/0112090].

\bibitem{Magueijo}
  J.~Magueijo and L.~Smolin,
  {\it Generalized Lorentz invariance with an invariant energy scale,}
  Phys.\ Rev.\  D {\bf 67} (2003) 044017
  [arXiv:gr-qc/0207085].

%\cite{KowalskiGlikman:2002we}
\bibitem{KowalskiGlikman:2002we}
  J.~Kowalski-Glikman and S.~Nowak,
  {\it Doubly special relativity theories as different bases of $\kappa$-Poincar\'e
  algebra,}
  Phys.\ Lett.\  B {\bf 539} (2002) 126
   [arXiv:hep-th/0203040].

%\cite{KowalskiGlikman:2002jr}
\bibitem{KowalskiGlikman:2002jr}
  J.~Kowalski-Glikman and S.~Nowak,
  {\it Non-commutative space-time of doubly special relativity theories,}
  Int.\ J.\ Mod.\ Phys.\  D {\bf 12} (2003) 299
  [arXiv:hep-th/0204245].

%\cite{Kosinski:1999ix}
\bibitem{Kosinski:1999ix}
  P.~Kosinski, J.~Lukierski and P.~Maslanka,
  {\it Local D = 4 field theory on $\kappa$-deformed Minkowski space,}
  Phys.\ Rev.\  D {\bf 62} (2000) 025004
  [arXiv:hep-th/9902037].

%\cite{Kosinski:1999dw}
\bibitem{Kosinski:1999dw}
  P.~Kosinski, J.~Lukierski and P.~Maslanka,
  {\it Local field theory on $\kappa$-Minkowski space, star products and
  noncommutative translations,}
  Czech.\ J.\ Phys.\  {\bf 50} (2000) 1283
  [arXiv:hep-th/0009120].

%\cite{AmelinoCamelia:2001fd}
\bibitem{AmelinoCamelia:2001fd}
  G.~Amelino-Camelia and M.~Arzano,
  {\it Coproduct and star product in field theories on Lie-algebra non-commutative
  space-times,}
  Phys.\ Rev.\  D {\bf 65} (2002) 084044
  [arXiv:hep-th/0105120].

%\cite{Daszkiewicz:2004xy}
\bibitem{Daszkiewicz:2004xy}
  M.~Daszkiewicz, K.~Imilkowska, J.~Kowalski-Glikman and S.~Nowak,
  {\it Scalar field theory on $\kappa$-Minkowski space-time and doubly special
  relativity,}
  Int.\ J.\ Mod.\ Phys.\  A {\bf 20} (2005) 4925
   [arXiv:hep-th/0410058].
  
%\cite{Dimitrijevic:2003wv}
\bibitem{Dimitrijevic:2003wv}
  M.~Dimitrijevic, L.~Jonke, L.~Moller, E.~Tsouchnika, J.~Wess and M.~Wohlgenannt,
 {\it Deformed field theory on $\kappa$-spacetime,}
  Eur.\ Phys.\ J.\  C {\bf 31} (2003) 129
  [arXiv:hep-th/0307149].
  
%\cite{Ghosh:2006cb}
\bibitem{Ghosh:2006cb}
  S.~Ghosh,
{\it A Lagrangian for DSR Particle and the Role of Noncommutativity,}
  Phys.\ Rev.\  {\bf D74}, 084019 (2006).
  [hep-th/0608206].
  
%\cite{Ghosh:2007ai}
\bibitem{Ghosh:2007ai}
  S.~Ghosh, P.~Pal,
 {\it Deformed Special Relativity and Deformed Symmetries in a Canonical Framework,}
  Phys.\ Rev.\  {\bf D75 } (2007)  105021.
  [hep-th/0702159].

\bibitem{Govindarajan:2009wt}
  T.~R.~Govindarajan, K.~S.~Gupta, E.~Harikumar, S.~Meljanac and D.~Meljanac,
  {\it Deformed Oscillator Algebras and QFT in $\kappa$-Minkowski Spacetime,}
  Phys. Rev. D {\bf 80} (2009) 025014, [arXiv:0903.2355 [hep-th]].
  
%\cite{Young:2007ag}
\bibitem{Young:2007ag}
  C.~A.~S.~Young and R.~Zegers,
 {\it Covariant particle statistics and intertwiners of the $\kappa$-deformed
  Poincare algebra,}
  Nucl.\ Phys.\  B {\bf 797} (2008) 537
  [arXiv:0711.2206 [hep-th]].
  %%CITATION = NUPHA,B797,537;%    
    
%\cite{Daszkiewicz:2007az}
\bibitem{Daszkiewicz:2007az}
  M.~Daszkiewicz, J.~Lukierski, M.~Woronowicz,
  {\it ``kappa-deformed statistics and classical fourmomentum addition law,}
  Mod.\ Phys.\ Lett.\  {\bf A23 } (2008)  653-665
  [hep-th/0703200].

%\cite{Arzano:2007ef}
\bibitem{Arzano:2007ef}
  M.~Arzano and A.~Marciano,
 {\it Fock space, quantum fields and $\kappa$-Poincar\'e symmetries,}
  Phys.\ Rev.\  D {\bf 76} (2007) 125005
  [arXiv:0707.1329 [hep-th]].
  %%CITATION = PHRVA,D76,125005;%%

%\cite{Daszkiewicz:2007ru}
\bibitem{Daszkiewicz:2007ru}
  M.~Daszkiewicz, J.~Lukierski and M.~Woronowicz,
 {\it Towards quantum noncommutative $\kappa$-deformed field theory,}
  Phys.\ Rev.\  D {\bf 77} (2008) 105007
  [arXiv:0708.1561].
  %%CITATION = PHRVA,D77,105007;%%

%\cite{Daszkiewicz:2008bm}
\bibitem{Daszkiewicz:2008bm}
  M.~Daszkiewicz, J.~Lukierski and M.~Woronowicz,
 {\it $\kappa$-deformed oscillators, the choice of star product and free
  $\kappa$-deformed quantum fields,}
  J.\ Phys.\ A  {\bf 42} (2009) 355201
  [arXiv:0807.1992 [hep-th]].
  %%CITATION = JPAGB,A42,355201;%%
  
%\cite{Arzano:2008bt}
\bibitem{Arzano:2008bt}
  M.~Arzano and D.~Benedetti,
 {\it Rainbow statistics,}
  Int.\ J.\ Mod.\ Phys.\  A {\bf 24} (2009) 4623
  [arXiv:0809.0889 [hep-th]].
  %%CITATION = IMPAE,A24,4623;%%
  
\bibitem{drinfeld} 
  V. G. Drinfel'd,
{\it Hopf algebras and the quantum Yang-Baxter equation}, 
Dokl. Akad. Nauk SSSR 283 (1985), no. 5, 1060-1064 (Russian);
translation in Sov. Math. Dokl. {\bf 32} (1985) 254.

\bibitem{drinfeldN}
V. G. Drinfel'd, {\it Quasi-Hopf algebras,}
Algebra i Analiz 1 (1989), no. 6, 114-148 (Russian);
translation in Leningrad Math. J. {\bf 1} (1990), no. 6, 1419-1457.
  
%\cite{Borowiec:2004xj}
\bibitem{Borowiec:2004xj}
  A.~Borowiec, J.~Lukierski and V.~N.~Tolstoy,
  {\it Jordanian quantum deformations of D = 4 anti-de-Sitter and Poincare
  superalgebras,}
  Eur.\ Phys.\ J.\  C {\bf 44} (2005) 139
  [arXiv:hep-th/0412131].%

%\cite{Borowiec:2006fc}
\bibitem{Borowiec:2006fc}
  A.~Borowiec, J.~Lukierski and V.~N.~Tolstoy,
  {\it Jordanian twist quantization of D=4 Lorentz and Poincare algebras and D=3
  contraction limit,}
  Eur.\ Phys.\ J.\  C {\bf 48}, 633 (2006)
  [arXiv:hep-th/0604146].
  %%CITATION = EPHJA,C48,633;%%
  
%\cite{Balachandran:2007vx}
\bibitem{Balachandran:2007vx}
  A.~P.~Balachandran, A.~Pinzul and B.~A.~Qureshi,
 {\it Twisted Poincare Invariant Quantum Field Theories,}
  Phys.\ Rev.\  D {\bf 77} (2008) 025021
  [arXiv:0708.1779 [hep-th]].
  
%\cite{Bu:2006dm}
\bibitem{Bu:2006dm}
  J.~G.~Bu, H.~C.~Kim, Y.~Lee, C.~H.~Vac and J.~H.~Yee,
  {\it $\kappa$-deformed Spacetime From Twist,}
  Phys.\ Lett.\  B {\bf 665} (2008) 95
  [arXiv:hep-th/0611175].

%\cite{Govindarajan:2008qa}
\bibitem{Govindarajan:2008qa}
  T.~R.~Govindarajan, K.~S.~Gupta, E.~Harikumar, S.~Meljanac and D.~Meljanac,
  {\it Twisted Statistics in $\kappa$-Minkowski Spacetime,}
  Phys.\ Rev.\  D {\bf 77} (2008) 105010, [arXiv:0802.1576 [hep-th]].

%\cite{Borowiec:2008uj}
\bibitem{Borowiec:2008uj}
  A.~Borowiec and A.~Pachol,
  {\it $\kappa$-Minkowski spacetime as the result of Jordanian twist deformation,}
  Phys.\ Rev.\  D {\bf 79} (2009) 045012
  [arXiv:0812.0576 [math-ph]].
  
  %\cite{Bu:2009tc}
\bibitem{Bu:2009tc}
  J.~G.~Bu, J.~H.~Yee and H.~C.~Kim,
  {\it Differential Structure on $\kappa$-Minkowski Spacetime Realized as Module of
  Twisted Weyl Algebra,}
  Phys.\ Lett.\  B {\bf 679} (2009) 486
  [arXiv:0903.0040 [hep-th]].

%\cite{Kim:2009jk}
\bibitem{Kim:2009jk}
  H.~C.~Kim, Y.~Lee, C.~Rim and J.~H.~Yee,
  {\it Scalar Field theory in $\kappa$-Minkowski spacetime from twist,}
  arXiv:0901.0049 [hep-th].

%\cite{Young:2008zm}
\bibitem{Young:2008zm}
  C.~A.~S.~Young and R.~Zegers,
  {\it On $\kappa$-deformation and triangular quasibialgebra structure,}
  Nucl.\ Phys.\  B {\bf 809} (2009) 439
  [arXiv:0807.2745 [hep-th]].
  %%CITATION = NUPHA,B809,439;%%
  
%\cite{Young:2008zg}
\bibitem{Young:2008zg}
  C.~A.~S.~Young and R.~Zegers,
 {\it Covariant particle exchange for $\kappa$-deformed theories in 1+1
  dimensions,}
  Nucl.\ Phys.\  B {\bf 804} (2008) 342
  [arXiv:0803.2659 [hep-th]].
  %%CITATION = NUPHA,B804,342;%%

%\cite{Freidel:2006gc}
\bibitem{Freidel:2006gc}
  L.~Freidel, J.~Kowalski-Glikman and S.~Nowak,
  {\it From noncommutative $\kappa$-Minkowski to Minkowski space-time,}
  Phys.\ Lett.\  B {\bf 648} (2007) 70
  [arXiv:hep-th/0612170].

%\cite{KowalskiGlikman:2009zu}
\bibitem{KowalskiGlikman:2009zu}
  J.~Kowalski-Glikman and A.~Walkus,
  {\it Star product and interacting fields on $\kappa$-Minkowski space,}
  Mod.\ Phys.\ Lett.\  A {\bf 24} (2009) 2243
  [arXiv:0904.4036 [hep-th]].

%\cite{Grosse:2005iz}
\bibitem{Grosse:2005iz}
  H.~Grosse and M.~Wohlgenannt,
  {\it On $\kappa$-deformation and UV/IR mixing,}
  Nucl.\ Phys.\  B {\bf 748} (2006) 473
  [arXiv:hep-th/0507030].

%\cite{Harikumar:2009wv}
\bibitem{Harikumar:2009wv}
  E.~Harikumar and M.~Sivakumar,
{\it Testable signatures of the $\kappa$-Minkowski Spacetime 
from the Hydrogen atom spectrum,}
  [arXiv:0910.5778].
  %%CITATION = ARXIV:0910.5778;%%

%\cite{Arzano:2009bd}
\bibitem{Arzano:2009bd}
  M.~Arzano, J.~Kowalski-Glikman and A.~Walkus,
 {\it A Bound on Planck-scale modifications of 
 the energy-momentum composition
  rule from atomic interferometry,}
  Europhys.\ Lett.\  {\bf 90} (2010) 30006
  [arXiv:0912.2712].
  %%CITATION = EULEE,90,30006;%%

%\cite{Borowiec:2009ty}
\bibitem{Borowiec:2009ty}
  A.~Borowiec, K.~S.~Gupta, S.~Meljanac and A.~Pachol,
{\it Constarints on the quantum gravity scale from $\kappa$-Minkowski
  spacetime,}
  Europhys.\ Lett.\  {\bf 92} (2010) 20006
  [arXiv:0912.3299 ].
  %%CITATION = EULEE,92,20006;%%
  
%\cite{Kosinski:1994br}
\bibitem{Kosinski:1994br}
  P.~Kosinski, J.~Lukierski, P.~Maslanka and J.~Sobczyk,
 {\it The Classical basis for $\kappa$ deformed Poincar\'e (super)algebra and the
  second $\kappa$ deformed supersymmetric Casimir,}
  Mod.\ Phys.\ Lett.\  A {\bf 10} (1995) 2599
  [arXiv:hep-th/9412114].
  %%CITATION = MPLAE,A10,2599;%%
 
%\cite{Borowiec:2009vb}
\bibitem{Borowiec:2009vb}
  A.~Borowiec and A.~Pachol,
{\it Classical basis for $\kappa$-Poincar\'e algebra and doubly special relativity
  theories,}
  J.\ Phys.\ A  {\bf 43} (2010) 045203
  [arXiv:0903.5251].
  %%CITATION = JPAGB,A43,045203;%%

%\cite{Meljanac:2006ui}
\bibitem{Meljanac:2006ui}
  S.~Meljanac and M.~Stojic,
 {\it New realizations of Lie algebra $\kappa$-deformed Euclidean space,}
  Eur.\ Phys.\ J.\  C {\bf 47} (2006) 531, [arXiv:hep-th/0605133].

%\cite{Meljanac:2007xb}
\bibitem{Meljanac:2007xb}
  S.~Meljanac, A.~Samsarov, M.~Stojic and K.~S.~Gupta,
  {\it $\kappa$-Minkowski space-time and the star product realizations,}
  Eur.\ Phys.\ J.\  C {\bf 53} (2008) 295, [arXiv:0705.2471 [hep-th]].
  
%\cite{KresicJuric:2007nh}
\bibitem{KresicJuric:2007nh}
  S.~Kresic-Juric, S.~Meljanac and M.~Stojic,
  {\it Covariant realizations of $\kappa$-deformed space,}
  Eur.\ Phys.\ J.\  C {\bf 51} (2007) 229, [arXiv:hep-th/0702215].
    
%\cite{Meljanac:2010ps}
\bibitem{Meljanac:2010ps}
  S.~Meljanac and A.~Samsarov,
{\it Scalar field theory on $\kappa$-Minkowski spacetime and translation and
  Lorentz invariance,}
  Int.\ J.\ Mod.\ Phys.\  A {\bf 26} (2011) 1439
  [arXiv:1007.3943].
  %%CITATION = IMPAE,A26,1439;%%
  
%\bibitem{GWul}
%\cite{Grosse:2005da}
\bibitem{Grosse:2005da}
  H.~Grosse and R.~Wulkenhaar,
  {\it Renormalisation of phi**4-theory on non-commutative R**4 to all orders,}
  Lett.\ Math.\ Phys.\  {\bf 71} (2005) 13 [arXiv:hep-th/0403232].
  %%CITATION = LMPHD,71,13;%%
%\cite{Grosse:2004yu}

\bibitem{Grosse:2004yu}
  H.~Grosse and R.~Wulkenhaar,
  {\it Renormalisation of phi**4 theory on noncommutative R**4 in the matrix
  base,}
  Commun.\ Math.\ Phys.\  {\bf 256} (2005) 305
  [arXiv:hep-th/0401128].
  %%CITATION = CMPHA,256,305;%%
    
%\cite{Horvat:2010km}
\bibitem{Horvat:2010km}
  R.~Horvat, J.~Trampetic,
 {\it Constraining noncommutative field theories with holography},
  JHEP {\bf 1101}, 112 (2011)
  [arXiv:1009.2933 [hep-ph]].  
  
  %\cite{Cohen:1998zx}
\bibitem{Cohen:1998zx}
  A.~G.~Cohen, D.~B.~Kaplan and A.~E.~Nelson,
 {\it Effective field theory, black holes, and the cosmological constant,}
  Phys.\ Rev.\ Lett.\  {\bf 82}, 4971 (1999)
  [arXiv:hep-th/9803132].
  
%\cite{Szabo:2009tn}
\bibitem{Szabo:2009tn}
  R.~J.~Szabo,
 {\it Quantum Gravity, Field Theory and Signatures of Noncommutative Spacetime,}
  Gen.\ Rel.\ Grav.\  {\bf 42 } (2010)  1-29
  [arXiv:0906.2913 [hep-th]].
  %
%\cite{Horvat:2011iv}
\bibitem{Horvat:2011iv}
  R.~Horvat, D.~Kekez, P.~Schupp, J.~Trampetic, J.~You,
 {\it Photon-neutrino interaction in theta-exact covariant noncommutative field theory,}
   Phys.\ Rev.\ D\ {\bf 84} (2011) 045004
  [arXiv:1103.3383 [hep-ph]].
  %%CITATION = PHRVA,D84,045004;%%

%\cite{arXiv:1109.2485}
\bibitem{arXiv:1109.2485}
  R.~Horvat, A.~Ilakovac, J.~Trampetic and J.~You,
 {\it On UV/IR mixing in noncommutative gauge field theories,}
  arXiv:1109.2485 [hep-th], to be published in JHEP.
  %%CITATION = ARXIV:1109.2485;%%

%\cite{arXiv:1111.4951}
\bibitem{arXiv:1111.4951}
  R.~Horvat, A.~Ilakovac, P.~Schupp, J.~Trampetic and J.~You,
 {\it Neutrino propagation in noncommutative spacetimes,}
  arXiv:1111.4951 [hep-th].
  %%CITATION = ARXIV:1111.4951;%
  
%\cite{KowalskiGlikman:2004qa}
\bibitem{KowalskiGlikman:2004qa}
  J.~Kowalski-Glikman,
 {\it Introduction to doubly special relativity,}
  Lect.\ Notes Phys.\  {\bf 669} (2005) 131
  [arXiv:hep-th/0405273].
  %%CITATION = LNPHA,669,131;%%
  
%\cite{AmelinoCamelia:2001me}
\bibitem{AmelinoCamelia:2001me}
  G.~Amelino-Camelia, J.~Lukierski and A.~Nowicki,
 {\it Absorption of TeV photons and $\kappa$ deformation of relativistic
  symmetries,}
  Czech.\ J.\ Phys.\  {\bf 51} (2001) 1247
  [arXiv:hep-th/0103227].
  %%CITATION = CZYPA,51,1247;%%

%\cite{Abel:2006wj}
\bibitem{Abel:2006wj}
  S.~A.~Abel, J.~Jaeckel, V.~V.~Khoze and A.~Ringwald,
 {\it Vacuum Birefringence as a Probe of Planck Scale Noncommutativity,}
  JHEP {\bf 0609} (2006) 074
  [arXiv:hep-ph/0607188].
  %%CITATION = JHEPA,0609,074;%%    

%\cite{Buric:2010wd}
\bibitem{Buric:2010wd}
  M.~Buric, D.~Latas, V.~Radovanovic and J.~Trampetic,
{\it Chiral fermions in noncommutative electrodynamics: renormalizability and
  dispersion,}
  Phys.\ Rev.\  D {\bf 83} (2011) 045023,
  arXiv:1009.4603 [hep-th].
  %%CITATION = ARXIV:1009.4603;%%
  
%\cite{Chang:2001bm}
%\bibitem{Chang:2001bm}
%L.~N.~Chang, D.~Minic, N.~Okamura and T.~Takeuchi,
%{\it The effect of the minimal length uncertainty relation on the density of
%states and the cosmological constant problem},
%Phys.\ Rev.\  D {\bf 65} (2002) 125028
%[arXiv:hep-th/0201017];
       %%CITATION = PHRVA,D65,125028;%%
%
%\cite{Benczik:2005bh}
%\bibitem{Benczik:2005bh}
%S.~Benczik, L.~N.~Chang, D.~Minic and T.~Takeuchi,
%{\it The hydrogen atom with minimal length},
%Phys.\ Rev.\  A {\bf 72} (2005) 012104
 % [arXiv:hep-th/0502222];
  %%CITATION = PHRVA,A72,012104;%%
%
%\bibitem{kmm}
%A. Kempf, G. Mangano and R. Mann,
%{\it Hilbert space representation of the minimal length uncertainty relation},
%Phys. Rev. D{\bf 52} (1995) 1108
%
%\cite{Borowiec:2010yw}
\bibitem{Borowiec:2010yw}
  A.~Borowiec, A.~Pachol,
 {\it $\kappa$-Minkowski spacetimes and DSR algebras: Fresh look and old problems,}
  SIGMA {\bf 6 } (2010)  086.
  [arXiv:1005.4429 [math-ph]] 
%  
%\cite{Maggiore:1993rv}
\bibitem{Maggiore:1993rv}
  M.~Maggiore,
  {\it A Generalized uncertainty principle in quantum gravity,}
  Phys.\ Lett.\  B {\bf 304} (1993) 65
  [arXiv:hep-th/9301067].
  %\cite{Maggiore:1993zu}
%
\bibitem{Maggiore:1993zu}
  M.~Maggiore,
  {\it Quantum Groups, Gravity And The Generalized Uncertainty Principle,}
  Phys.\ Rev.\  D {\bf 49} (1994) 5182
  [arXiv:hep-th/9305163].
  %%CITATION = PHRVA,D49,5182;%%
  
%\cite{Giddings:2007bw}
\bibitem{Giddings:2007bw}
  S.~B.~Giddings, D.~J.~Gross and A.~Maharana,
{\it Gravitational effects in ultrahigh-energy string scattering,}
  Phys.\ Rev.\  D {\bf 77} (2008) 046001
  [arXiv:0705.1816 [hep-th]].
  %%CITATION = PHRVA,D77,046001;%%  
 
%\cite{majid1}
\bibitem{majid1}
  S. Majid,
  {\it Foundations of Quantum Group Theory}, Cambridge University 
  Press, 1995.
  
%\cite{Oeckl:2000eg}
\bibitem{Oeckl:2000eg}
  R.~Oeckl,
  {\it Untwisting noncommutative R**d and the equivalence of quantum field
  theories,}
  Nucl.\ Phys.\  B {\bf 581} (2000) 559
  [arXiv:hep-th/0003018].
  %%CITATION = NUPHA,B581,559;%%
  
%\cite{wess} 
%\cite{Aschieri:2005zs}
\bibitem{Aschieri:2005zs}
  P.~Aschieri, M.~Dimitrijevic, F.~Meyer and J.~Wess,
  {\it Noncommutative geometry and gravity,}
  Class.\ Quant.\ Grav.\  {\bf 23} (2006) 1883
  [arXiv:hep-th/0510059].
  %%CITATION = CQGRD,23,1883;%%

%\cite{wess1}  
%\cite{Aschieri:2005yw}
\bibitem{Aschieri:2005yw}
  P.~Aschieri, C.~Blohmann, M.~Dimitrijevic, F.~Meyer, P.~Schupp and J.~Wess,
  {\it A gravity theory on noncommutative spaces,}
  Class.\ Quant.\ Grav.\  {\bf 22} (2005) 3511
  [arXiv:hep-th/0504183].
  %%CITATION = CQGRD,22,3511;%% 
  
%\cite{Balachandran:2005eb}
\bibitem{Balachandran:2005eb}
  A.~P.~Balachandran, G.~Mangano, A.~Pinzul and S.~Vaidya,
  {\it Spin and statistics on the Groenwald-Moyal plane: Pauli-forbidden  levels
  and transitions,}
  Int.\ J.\ Mod.\ Phys.\  A {\bf 21} (2006) 3111
  [arXiv:hep-th/0508002].
  %%CITATION = IMPAE,A21,3111;
  
%\cite{Balachandran:2006pi}
\bibitem{Balachandran:2006pi}
  A.~P.~Balachandran, T.~R.~Govindarajan, G.~Mangano, A.~Pinzul, 
  B.~A.~Qureshi and S.~Vaidya,
 {\it Statistics and UV-IR mixing with twisted Poincare invariance,}
  Phys.\ Rev.\  D {\bf 75} (2007) 045009
  [arXiv:hep-th/0608179].

%\cite{Balachandran:2010xk}
\bibitem{Balachandran:2010xk}
  A.~P.~Balachandran, A.~Joseph and P.~Padmanabhan,
  {\it Non-Pauli Transitions From Spacetime Noncommutativity},
  Phys.\ Rev.\ Lett.\  {\bf 105} (2010) 051601
  [arXiv:1003.2250 [hep-th]].

%\cite{Balachandran:2010wq}
\bibitem{Balachandran:2010wq}
  A.~P.~Balachandran and P.~Padmanabhan,
  {\it Non-Pauli Effects from Noncommutative Spacetimes,}
  JHEP {\bf 1012} (2010) 001
  [arXiv:1006.1185 [hep-th]].
 
\bibitem{vgdrinfeld}
 V. G. Drinfeld, 
 {\it Quantum groups,} 
 Proceedings of the ICM, Rhode Island USA, 1987.

\bibitem{faddeev}
L. D. Faddeev, N. Yu. Reshetikhin and L. A. Takhtajan, {\it Quantization of
Lie groups and Lie algebras}, Algebra i Analiz. {\bf 1},
(1989). English transl. in Leningrad Math. J.

\bibitem{Rmatrixmajid}
 S. Majid, {\it Quasitriangular Hopf algebras and Yang-Baxter equations},
 Int. J. Mod. Phys. A {\bf 5} (1990) 1.  

%\cite{KowalskiGlikman:2002ft}
\bibitem{KowalskiGlikman:2002ft}
  J.~Kowalski-Glikman,
 {\it De sitter space as an arena for doubly special relativity,}
  Phys.\ Lett.\  B {\bf 547} (2002) 291
  [arXiv:hep-th/0207279].
  %%CITATION = PHLTA,B547,291;%%

%\cite{Lukierski:2002wf}
\bibitem{Lukierski:2002wf}
  J.~Lukierski and A.~Nowicki,
 {\it Nonlinear and quantum origin of doubly infinite family of modified addition
  laws for four momenta,}
  Czech.\ J.\ Phys.\  {\bf 52} (2002) 1261
  [arXiv:hep-th/0209017].
  %%CITATION = CZYPA,52,1261;%%
  
%\cite{Blaschke:2009aw}
\bibitem{Blaschke:2009aw}
  D.~N.~Blaschke, H.~Grosse, E.~Kronberger, M.~Schweda and M.~Wohlgenannt,
 {\it Loop Calculations for the Non-Commutative U(*)(1) Gauge Field Model with
  Oscillator Term,}
  Eur.\ Phys.\ J.\  C {\bf 67} (2010) 575
  [arXiv:0912.3642].
  %%CITATION = EPHJA,C67,575;%%
  
%\cite{Agostini:2006nc}
\bibitem{Agostini:2006nc}
  A.~Agostini, G.~Amelino-Camelia, M.~Arzano, A.~Marciano and R.~A.~Tacchi,
 {\it Generalizing the Noether theorem for Hopf-algebra spacetime symmetries,}
  Mod.\ Phys.\ Lett.\  A {\bf 22} (2007) 1779
  [arXiv:hep-th/0607221].
  %%CITATION = MPLAE,A22,1779;%%

%\cite{Kosinski:2003xx}
\bibitem{Kosinski:2003xx}
  P.~Kosinski, P.~Maslanka, J.~Lukierski and A.~Sitarz,
 {\it Generalized $\kappa$ deformations and deformed relativistic scalar fields on
  noncommutative Minkowski space,}
 Published in *Mexico City 2002, 
 Topics in mathematical physics, general relativity and cosmology* 255-277;
  arXiv:hep-th/0307038.
  %%CITATION = HEP-TH/0307038;%%
  
  \bibitem{Casalbuoni} 
  Roberto Casalbuoni,
  {\it Advanced Quantum Field Theory},
  Dipartimento di Fisica, Lezioni date all'Universita' di
  Firenze nell'a.a. 2004/2005.
  
%\cite{Balachandran:2007sh}
%\bibitem{Balachandran:2007sh}
%  A.~P.~Balachandran, A.~G.~Martins and P.~Teotonio-Sobrinho,
% {\it Discrete time evolution and energy nonconservation in noncommutative
%  physics,}
%  JHEP {\bf 0705} (2007) 066
%  [arXiv:hep-th/0702076].  

\end{thebibliography}
\end{document}